# Signatures of Chiral Superconductivity in Rhombohedral Graphene

Tonghang Han[1]†, Zhengguang Lu[1,2]†, Zach Hadjri[1]†, Lihan Shi[1], Zhenghan Wu[1], Wei Xu[1], Yuxuan Yao[1], Armel A. Cotten[3], Omid Sharifi Sedeh[3], Henok Weldeyesus[3], Jixiang Yang[1], Junseok Seo[1], Shenyong Ye[1], Muyang Zhou[1], Haoyang Liu[2], Gang Shi[2], Zhenqi Hua[2], Kenji Watanabe[4], Takashi Taniguchi[5], Peng Xiong[2], Dominik M. Zumbühl[3], Liang Fu[1], Long Ju[1]*

[1]Department of Physics, Massachusetts Institute of Technology, Cambridge, MA, 02139, USA.

[2]Department of Physics, Florida State University, Tallahassee, FL, 32306, USA

[3]Department of Physics, University of Basel, 4056, Basel, Switzerland

[4]Research Center for Electronic and Optical Materials, National Institute for Materials Science, 1-1 Namiki, Tsukuba 305-0044, Japan

[5]Research Center for Materials Nanoarchitectonics, National Institute for Materials Science, 1-1 Namiki, Tsukuba 305-0044, Japan

*Corresponding author. Email: longju@mit.edu †These authors contributed equally to this work.

**Chiral superconductors are unconventional superconducting states that break time reversal symmetry spontaneously and typically feature Cooper pairing at non-zero angular momentum. Such states may host Majorana fermions and provide an important platform for topological physics research and fault-tolerant quantum computing[1–7]. Despite intensive search and prolonged studies of several candidate systems[8–27], chiral superconductivity has remained elusive so far. Here we report the discovery of robust unconventional superconductivity in rhombohedral tetra- and penta-layer graphene in the absence of moiré superlattice effects. We observed two superconducting states in the gate-induced flat conduction bands with $T_c$ up to 300 mK and charge density $n_e$ as low as $2.4*10^{11}$ cm$^{-2}$ in three tetralayer and two pentalayer devices. Spontaneous time-reversal-symmetry-breaking (TRSB) due to electron's orbital motion is found, and several observations indicate the chiral nature of these superconducting states, including: 1. In the superconducting state, $R_{xx}$ shows magnetic hysteresis in varying out-of-plane magnetic field $B_\perp$, which is absent from all other superconductors; 2. the superconducting states are immune to in-plane magnetic field and are developed within a spin- and valley-polarized quarter-metal phase; 3. the normal states show anomalous Hall signals at zero magnetic field and magnetic hysteresis. We also observed a critical $B_\perp$ of up to 1.4 Tesla, higher than any graphene superconductivity reported so far and indicates a strong-coupling superconductivity close to the BCS-BEC crossover[28]. Our observations establish a pure carbon material for the study of topological superconductivity, and pave the way to explore Majorana modes and topological quantum computing.**

Topological superconductivity has been conceived as new quantum states of matter, which host exotic quasiparticles that have great potential applications in quantum computing[1,2,4–7]. Chiral superconductors could host topological superconductivity with time-reversal-symmetry-breaking (TRSB) and magnetic hysteresis[2–4,6,29]. Several candidates of chiral superconductors have been investigated through a variety of experimental techniques since three decades ago[8–25,27]. Although signatures that are compatible with chiral superconductivity have been identified, most recent experimental reports suggest alternative pictures. For example, $UTe_2$ and $Sr_2RuO_4$ have been shown to have single-component order parameters that is incompatible with chiral superconductivity[17,30,31], and alternative origins of the observed TRSB were suggested[32]. In all these superconductors, there has been no evidence of anomalous Hall effect or magnetic hysteresis in their charge transport, making chiral superconductivity an elusive goal to be realized.

Graphene-based two-dimensional material heterostructures have emerged as a new playground for superconductivity with unconventional ingredients. By introducing the moiré superlattice between adjacent graphene layers[33–35], or between graphene and hBN[36], superconducting and correlated insulating states have been observed, reminiscent of the phase diagram of high-$T_c$ superconductors. More recently, it was shown that crystalline graphene in the rhombohedral stacking order could also exhibit superconductivity in the absence of moiré effects[37–45]. Rhombohedral stacked multilayer graphene hosts gate-tunable flat bands which drastically promotes correlation effects[46,47]. As shown in Fig. 1b, the conduction band in tetra-layer graphene becomes most flat when a gate-induced interlayer potential difference (between the top-most and bottom-most graphene layers) $\Delta$ = 90 meV, based on our tight-binding calculation (see Methods). Similar scenario happens in pentalayer graphene. As a result, various ground states with broken spin and/or valley symmetries due to the exchange interactions[48–51] have been observed. Such states with tunable fermi-surface topology and various spin/valley characters provide a fertile ground to search for unconventional superconductivity[52,53], including chiral superconductivity. Especially, interaction-induced valley polarization results in TRSB due to the chirality of electron motion, while the valley-dependent pseudo-spin winding[46,47,54] and angular-momentum[55,56] might facilitate high-angular-momentum pairing between electrons. The search of superconductivities in rhombohedral graphene, however, has been limited to three layers[37,43,44] so far, and the potential of unconventional superconductivity in this system is yet to be fully explored.

Here we report the DC transport study of rhombohedral stacked tetra- and penta-layer graphene devices. We observed superconductivity on the electron-doped side with the highest transition temperature of 300 mK. We measured three tetralayer and two pentalayer devices: Device T1 is tetra-layer graphene with electrons close to $WSe_2$, Device T2 is tetra-layer graphene with electrons away from $WSe_2$, Device T3 is bare tetra-layer graphene without $WSe_2$, Device P1 and P2 are bare penta-layer. All five devices show two unconventional superconducting states, in the absence of a detectable moiré superlattice. Several observations indicate TRSB and valley polarization in the observed superconducting states, most notably magnetic hysteresis in both the superconducting state and its corresponding normal state. These superconducting states persist to an out-of-plane magnetic field $B_\perp$ up to 1.4 Tesla – indicating a superconducting coherence length close to the inter-electron distance, and the underlying strongly coupled superconductivity picture[28,57]. We will focus on Device T3 and P1 in the main text, since there is no $WSe_2$ in them and the discussion is simpler. The data from Device T1, T2 and P2 are included as Extended Data Figures, where the influence of $WSe_2$ will also be discussed.

**Phase Diagram Showing Superconductivity**

Figure 1c shows the longitudinal resistance $R_{xx}$ map at the nominal base temperature of 7 mK at the mixing chamber, when Device T3 is electron-doped in the flat conduction band. At around $D/\varepsilon_0$ = 1.1 V/nm, three regions show vanishing resistances, as pointed by the arrows. Similar phase diagram is observed in Device P1 as shown in Fig. 1d, featuring three regions of vanishing resistances alike those in Device T3. We note that SC3 in Device P1 is not well-developed, while in Device P2 it is well-developed (see Extended Data Fig. 10). We name these three regions as SC1-3, since they are all superconducting states (see more data in Extended Data Figs. 4&9). Figure 1e&f show the temperature dependence of $R_{xx}$ in SC1-SC3. All three states show a transition to zero $R_{xx}$ as the temperature is lowered. The transition temperature reaches ~300 mK for SC1 in Device T3, highest among all superconducting states in Device T3 and P1. There is another superconducting SC4 state at high-electron-doping observed in Device T2 that is phenomenologically different from the superconductivity shown in Fig. 1c&d, especially SC1&2 (see Extended Data Figure 1).

SC1-3 reside at $n_e < 10^{12}$ cm$^{-2}$, corresponding to all the electrons located in the flat band bottom as shown in Fig. 1b at $\Delta$ = 90 mV, assuming the electrons are of the same spin and valley characters like in a quarter-metal. At the same time, SC1-3 are neighbored by a highly resistive region at lower densities, which is also reminiscent of the highly resistive region in tetra- to hexa-layer rhombohedral graphene/hBN moiré superlattices[58–60]. These observations of SC1-3 are in line with the expectation of strong electron correlation effects happening in the flat conduction band at intermediate $D$, as that is shown in Fig. 1b.

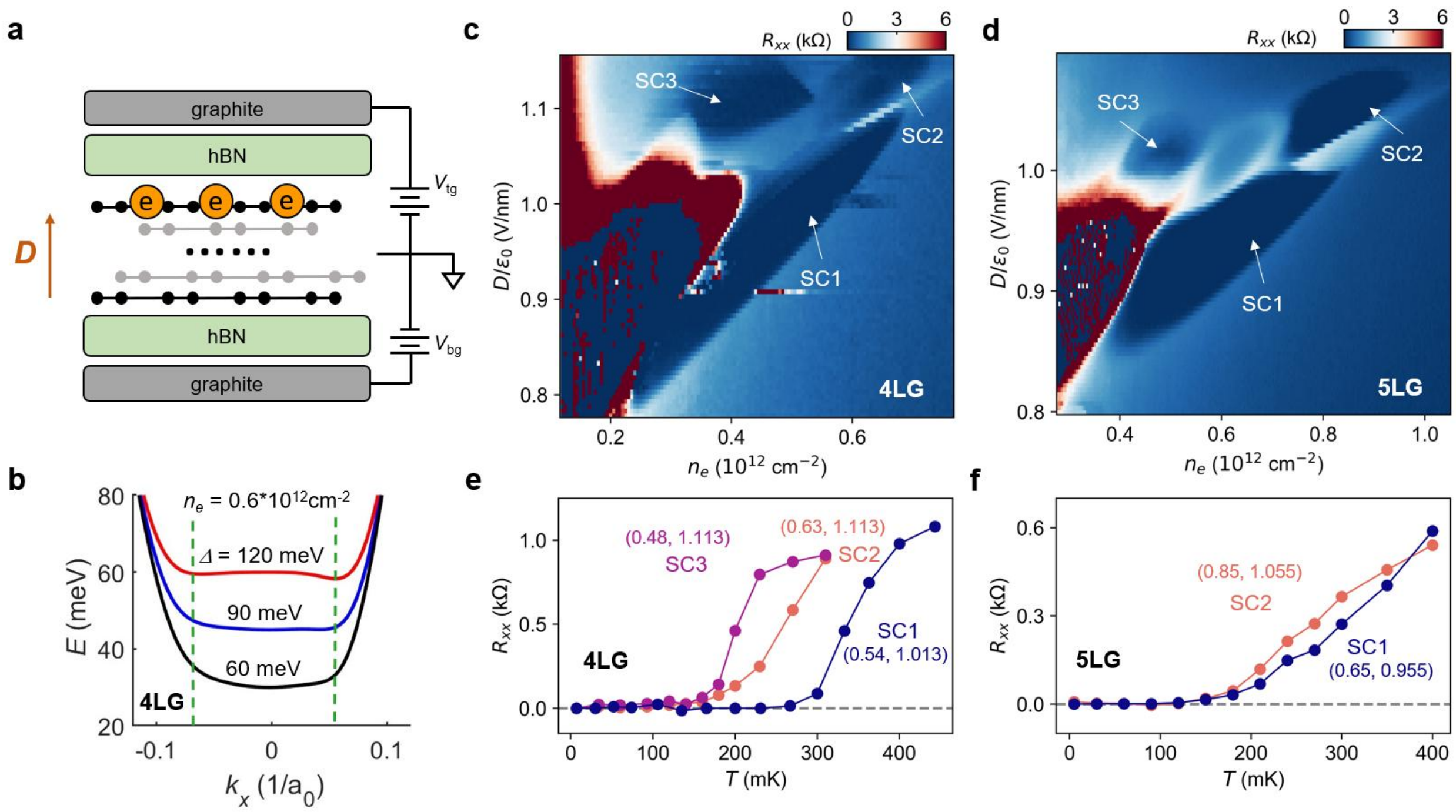

***Figure 1. Superconductivity in the flat bands of rhombohedral tetralayer graphene Device T3 and pentalayer graphene Device P1. a***, *Illustration of the device structure, in which the tetra- and penta-layer graphene form large twist angle with hBN to avoid the moiré superlattice effect.* ***b***, *The dispersion of conduction band in tetralayer graphene under varying potential difference between top and bottom layer $\Delta$, featuring a flat band bottom enclosing a charge density $n_e$ of $0.6*10^{12}$ cm$^{-2}$ per valley per spin at $\Delta$ = 90 meV.* ***c&d***, *Four-terminal resistance $R_{xx}$ as a function of $n_e$ and gate displacement field $D/\varepsilon_0$ taken at zero magnetic field and base temperature (7 mK at the mixing chamber) in tetra- and penta-layer graphene, respectively. Three regions show zero $R_{xx}$ (labeled as 'SC1-3' respectively) and superconductivity.* ***e&f***,

*Temperature dependence of the superconducting states in tetra- and penta-layer graphene, respectively. The SC1-3 curves are taken at the labeled (n, $D/\varepsilon_0$) in the units of ($10^{12}$ $cm^{-2}$, V/nm), respectively.*

**Neighboring Spin- and Valley-Polarized Quarter-Metal State**

To understand the superconductivities shown in Fig. 1 better, we first characterize the neighboring metallic states. We use the tetralayer Device T3 as an example, while the observations in the pentalayer Device P1 is similar (see Extended Data Figure 9). Figure 2a&b show the $R_{xx}$ and $R_{xy}$ maps taken at $B_\perp = 0.1$ T and the base temperature, in which the SC1 and SC2 regions can be clearly seen with vanishing values in both maps. SC3 is no longer visible in these maps, indicating an out-of-plane critical magnetic field less than 0.1 T. Furthermore, Fig. 2c shows the magnetic field scans taken in the states indicated by the purple and blue dots in Fig. 2b, respectively, revealing hysteretic loops in $R_{xx}$ and $R_{xy}$. Fig. 2d shows the $R_{xx}$ map taken at $B_\perp = 1$ T. The region neighboring the high-density boundary of SC1 and the low-$D$ boundary of SC2 shows clear quantum oscillations with a period that corresponds to that of a quarter-metal[48,50,51]. We did not observe quantum oscillations in the regions of SC1&2, possibly due to the extremely large effective mass and small cyclotron gap corresponding to the flat electron band in these regions (see Extended Data Figure 12).

The anomalous Hall signals and magnetic hysteresis shown in Fig. 2c clearly indicates a spontaneous valley polarization and TRS-breaking. Together with the quantum oscillation data in Fig. 2d, we conclude that SC1 and SC2 are neighbored by spin- and valley-polarized quarter-metals. The TRS is broken at the orbital level in these quarter-metal states, and the system spontaneously chooses a chirality in its electron transport at zero magnetic field due to the valley-polarization.

After establishing that spin- and valley-polarized quarter-metals are neighboring SC1 and SC2, we proceed to explore the evolution of the three states in magnetic field. Fig. 2e&f show the $R_{xx}$ and $R_{xy}$ taken along the dashed lines in Fig. 2a&b as a function of $B_\perp$. At this $D$, SC1 can persist to ~ 0.6 T before the $R_{xx}$ value starts to deviate from zero. The phase boundary between SC1 and the valley-polarized quarter-metal remains at the same $n_e$ as $B_\perp$ is increased. The left boundary even expands to lower density from $B_\perp = 0$ to 0.4 T, meaning that states in a small range of $n_e$ become superconducting only under a non-zero $B_\perp$.

The critical magnetic field of > 0.6 T in tetralayer graphene is unusually high for graphene superconductivity, and the corresponding value in the pentalayer device can even reach 1.4 T. We will discuss them in detail in Fig. 5. For now, we focus on the competition between SC1 and the neighboring states. If SC1 has zero orbital magnetization (or non-zero but smaller than that of the spin- and valley-polarized QM), the range of SC1 will shrink upon the application of $B_\perp$, since the energy of the QM will be lowered more than that of SC1 will be[37,48]. The observation of SC1 holding against the neighboring quarter-metal and even expanding implies the valley-polarization and orbital magnetic nature of SC1.

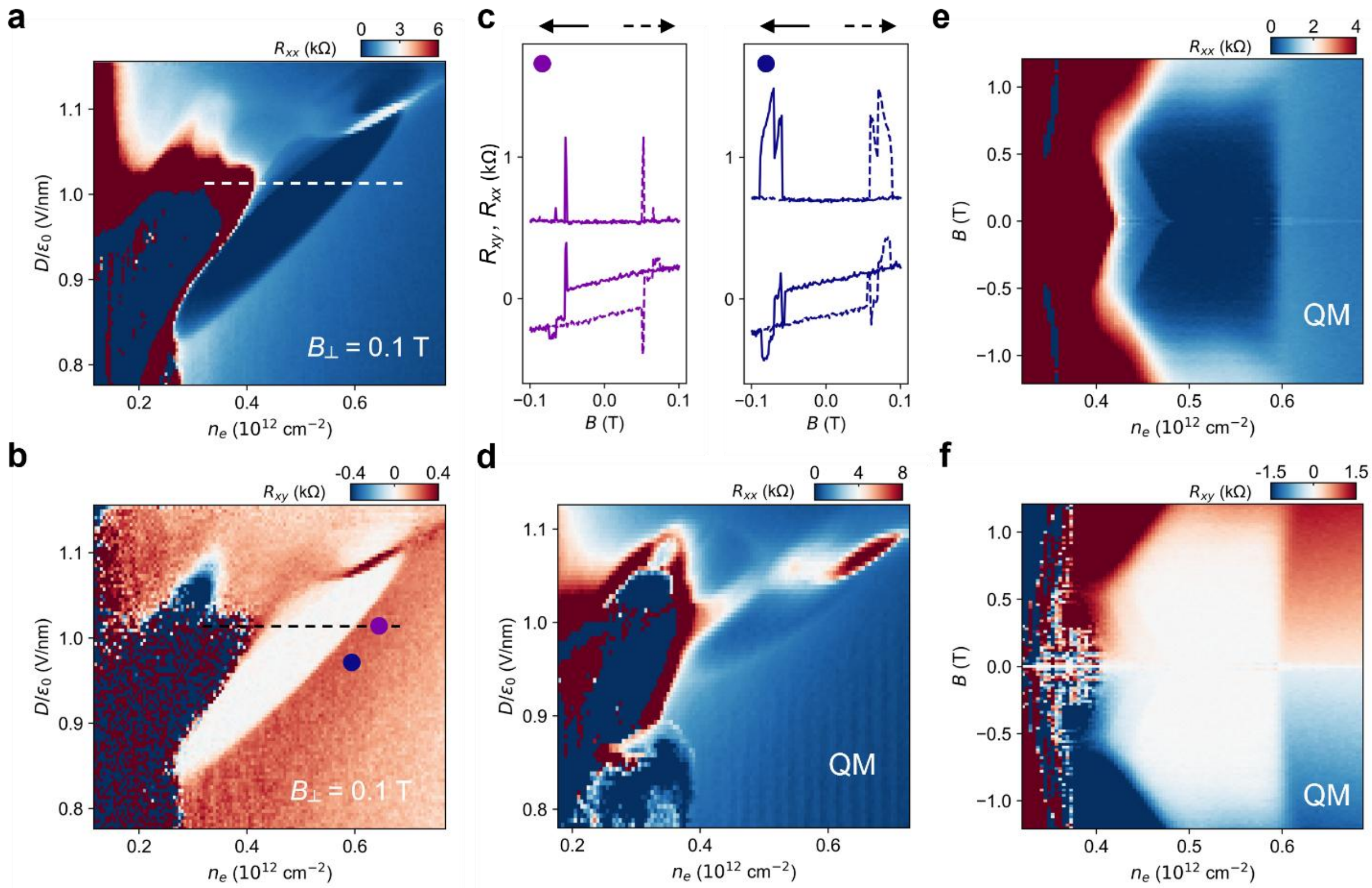

*Figure 2. TRSB and valley polarization in the neighboring states in the tetralayer Device T3.* ***a&b****,* $R_{xx}$ *and* $R_{xy}$ *maps at 0.1 T and base temperature (7 mK at the mixing chamber), extracted by symmetrizing and anti-symmetrizing the data taken at* $B_\perp = \pm 0.1$ *T. In* ***b****, SC1 (SC2) is surrounded (neighbored) by states that show anomalous Hall signals. The value of normal Hall signals at the same* $n_e$ *can be seen in the high-D part of the map.* ***c****,* $R_{xy}$ *and* $R_{xx}$ *during forward (dashed curves) and backward (solid curves) scans of* $B_\perp$ *at the purple and blue dots in* ***b****, respectively. The magnetic hysteresis and anomalous Hall signal indicate valley polarization.* ***d****,* $R_{xx}$ *map taken at* $B_\perp = 1$ *T. The period of quantum oscillations indicates a quarter metal (as labeled by 'QM') that neighbors SC1. Together with the data in* ***c****, this neighboring state to SC1 is a spin- and valley-polarized quarter-metal.* ***e&f****,* $R_{xx}$ *and* $R_{xy}$ *as a function of* $n_e$ *and* $B_\perp$ *along the dashed line in* ***a&b*** *(*$D/\varepsilon_0 = 1.013$ *V/nm), respectively. The phase boundary between the QM and SC1 remains at the same* $n_e$*, indicating the orbital magnetization is continuous across the boundary. The left boundary of SC1 (indicated by zero* $R_{xx}$ *and* $R_{xy}$*) even expands in magnetic field, confirming its orbital magnetic nature.*

### Spin- and Valley-Polarization in the Superconducting States

Knowing the spin- and valley-polarized QM nature of the neighboring metallic states, next we directly probe the spin and valley symmetry in SC1&2. Here we use data from the pentalayer Device P1 as an example. Figure 3a-d show the $n_e$-$D$ maps of $R_{xx}$ at in-plane magnetic field $B_{//} = 0$, 1, 3, 5 T, respectively (an out-of-plane magnetic field $B_\perp = 0.2$ T is applied to prevent the random fluctuation of $R_{xx}$, see Methods for discussion). Under an in-plane magnetic field, the zero-resistance of superconducting state survives in most of the SC1 region that was shown at zero magnetic field. Similar observation is made for SC2, as shown in Fig. 3e-h.

The robust superconductivity in a large in-plane magnetic field indicates the spin-polarized nature of SC1 and SC2. The Pauli-limit-Violation-Ratio (PVR) is already ~15 for SC1 at 5 T (see Methods), and the true PVR is likely much larger than 15, should we increase the magnetic field to even higher values to test. The lower limits of the in-plane critical field we observed is larger than in the spin-polarized superconductivity in bilayer graphene[45]. The spin-polarization of SC1 and SC2 indicates the connection between these superconducting states and the neighboring QM.

Figure 3i&j show the $R_{xx}$ under scanned $B_{\perp}$ in SC1 and SC2, respectively. Surprisingly, clear hysteresis between two zero-resistance states is observed for both superconductivities. A non-zero-resistance peak appears during the scanning, the magnetic field at which shows a hysteresis between forward and backward scans. Furthermore, such non-zero-resistance peak between zero-resistance states and the hysteresis behavior are observed even when a large in-plane magnetic field $B_{//}$ is applied, as shown in Fig. 3k&l.

The magnetic hysteresis of resistance in a superconducting state is highly unusual and distinct from all other superconductors: ferromagnetic superconductors show magnetic hysteresis in their optical responses but not in resistance directly[61]; magnetic hysteresis in resistance due to vortex-array-melting happen between the superconducting and metallic states[62], rather than between two superconducting states. The observations in Fig. 3i-l suggest the orbital magnetic nature of SC1 and SC2. This is illustrated in Fig. 3m: the two zero-resistance states at large $B_{\perp}$ field correspond to a single (and opposite-) valley-polarized domain between the voltage contacts in the device, while the non-zero resistance during scanning happens when a domain wall separates opposite-valley-polarized domains. This domain is expected to be resistive, as the tunneling of Cooper pair through it does not conserve momentum. The domain is flipped due to the coupling of valley-orbital-magnetization and the out-of-plane magnetic field —a mechanism similar to that induces the hysteresis shown in Fig. 2c. The possibility of domain flipping due to coupling to the spin magnetization is ruled out in two ways: 1. in Fig. 3k&l, the spin is always locked to the in-plane direction during scanning due to the much larger $B_{//}$ field than $B_{\perp}$ field; 2. when the valley polarization is fixed by a $B_{\perp}$ field, scanning the $B_{//}$ field in a large range does not induce any non-zero-resistance state or hysteresis, as shown in Fig. 3n&o. These observations strongly suggest the similarity and connection between the spin-valley-polarized QM and SC1&2.

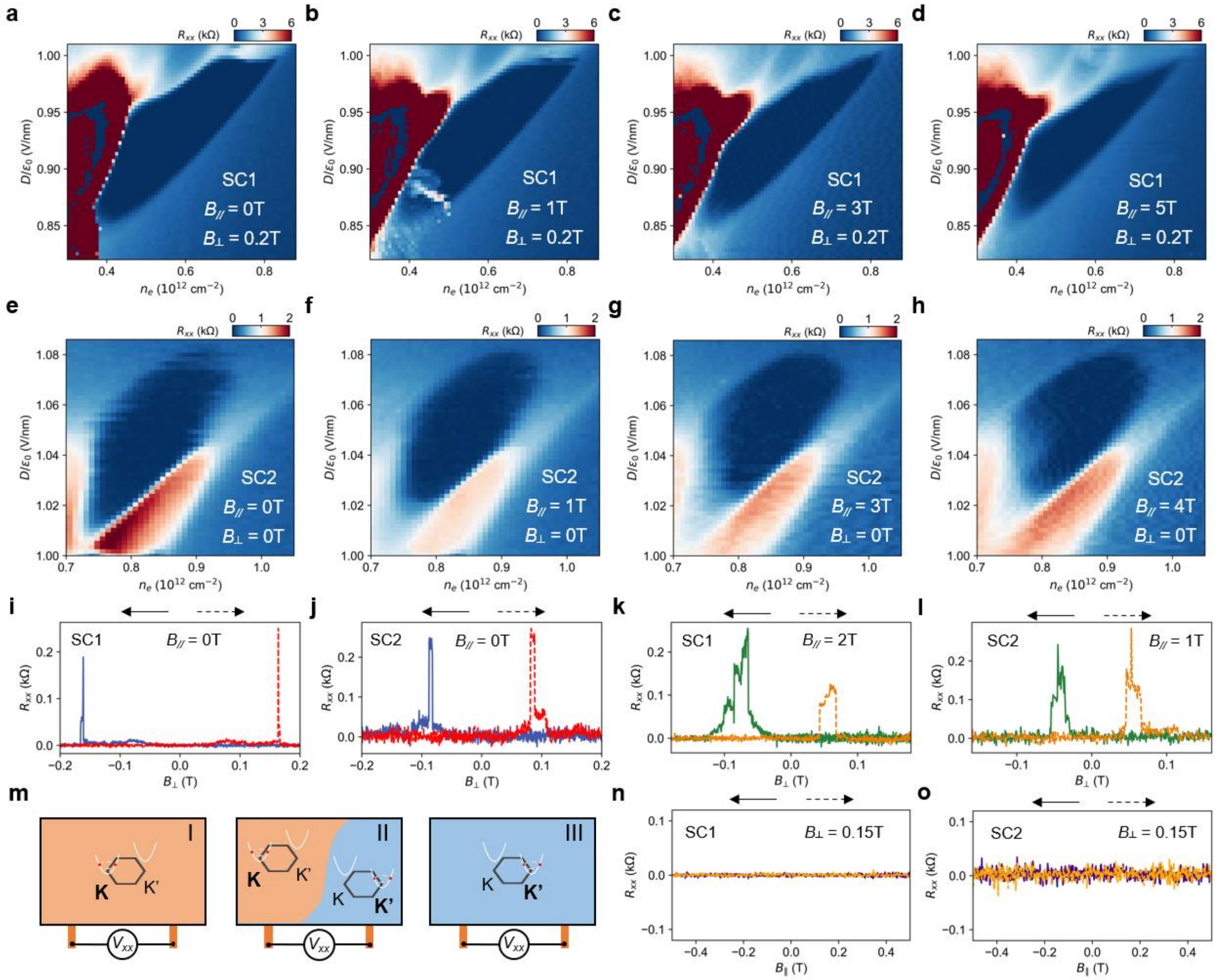

***Figure 3. Spin and valley polarization in the superconducting states in the pentalayer Device P1. a-d**, $n_e$-D maps of $R_{xx}$ in SC1 at in-plane magnetic field $B_{//}$ = 0, 1, 3, 5 T, respectively (an out-of-plane magnetic field $B_\perp$= 0.2 T is applied to prevent the random fluctuation of $R_{xx}$). **e-h**, $n_e$-D maps of $R_{xx}$ in SC2 at in-plane magnetic field $B_{//}$ = 0, 1, 3, 4 T, respectively with an out-of-plane magnetic field $B_\perp$= 0. **i&j**, $R_{xx}$ during forward (dashed curves) and backward (solid curves) scans of $B_\perp$ at the ($D/\varepsilon_0$ = 0.96 V/nm, $n_e$ = 0.7*$10^{12}$cm$^{-2}$) and ($D/\varepsilon_0$ = 1.05 V/nm, $n_e$ = 0.85*$10^{12}$cm$^{-2}$) in **a** and **d**, respectively. Clearly hysteresis between two zero-resistance states can be seen in both cases, indicating a ferro-magnet-like behavior of the superconductors. **k&l**, $R_{xx}$ during forward and backward scans of $B_\perp$ at the same states as in **i** and **j**, with $B_{//}$ = 2 T and 1 T applied, respectively. Similar hysteresis as in **i&j** can be seen, although the spin is fixed by the in-plane magnetic field. **m**, Illustration of the three states during the magnetic hysteresis scans in **i-l**, where state I and III correspond to uniform valley-polarized domains and zero-resistance states, and state II corresponds to a domain wall between oppositely valley-polarized domains between the voltage contacts and non-zero-resistance states. **n&o**, $R_{xx}$ during forward and backward scans of $B_\perp$ at the same states as in **i** and **j**, with $B_\perp$ = 0.15 T applied. No hysteresis is observed in either case, in stark contrast to **i-l**.*

**Temperature-Dependent Phase Evolution**

Another approach to understand the SC1&2 as well as their relation with the neighboring QM is to explore the corresponding normal states. Here we focus on data from the tetralayer Device T3 for the most complete characterization. The behaviors in other devices are qualitatively the same (see Extended Data Figures 9 for example). Figure 4a&b show the symmetrized $R_{xx}$ and anti-symmetrized $R_{xy}$ maps respectively at $B_\perp$ = 0.1 T and $T$ = 480 mK. The zero resistances in both SC1 and SC2 are replaced by values that are around 1-2 kΩ. In the Hall resistance map Fig. 4b, anomalous Hall signals of ~ 100 Ω are distributed in a region that overlaps with the SC1 and SC2 regions (outlined by the dashed oval-shaped curves). These anomalous Hall signals are confirmed by Fig. 4c, where the $R_{xy}$ at scanned magnetic fields are shown for representative $n_e$-$D$ combinations both within SC1 and in surrounding states (corresponding to the five symbols in Fig. 4b). Such magnetic hysteresis persists to 7 mK while $R_{xy}$ is zero in SC1 except for at the coercive fields, as shown in Fig. 4d. Figure 4e shows the evolution of $R_{xy}$ hysteresis as a function of temperature at the star position.

These observations suggest that the TRSB and valley-polarization already exist in the normal states of superconducting SC1 and SC2 states. To our knowledge, this is the first time that an anomalous Hall signal at zero magnetic field and magnetic hysteresis behavior are observed in the normal state of a superconductor, except for in hybrid systems where superconductivity and ferromagnetism co-exist[61,63–65]. These features are inherited by the electrons when they become superconducting at below the transition temperature. The Hall angles in these anomalous Hall states are quite large, corresponding to $\tan\theta_{\mathrm{H}} = \frac{R_{xy}}{R_{xx}} \sim$ up to 0.1, which is typical for quarter-metal states in crystalline rhombohedral graphene devices[48–51].

We note that there is a clear boundary intercepting the SC1 region in Fig. 4a, which corresponds to a sudden change of $R_{xx}$. This boundary is highlighted by the orange dashed curve in Fig. 4f. At a specific displacement field ($D/\varepsilon_0$= 0.923 V/nm for example, as shown in Fig. 4g), this phase boundary and kink in $R_{xx}$ gradually shift to lower $n_e$ during cooling down. At ~250 mK, the SC1 dome starts to develop in the region that is on the higher-density-side of this phase boundary. Figure 4h shows line-cuts at varying temperatures that highlight the kink and its temperature-dependent evolution.

By performing quantum oscillation measurements, we determine the higher-density-side of this boundary to be the spin- and valley-polarized quarter metal. It is hard to determine the Fermi surface topology of the lower-density-side due to the lack of clear quantum oscillations (we thus name it 'undetermined metal' or 'UM'), while one possibility is a metal state with annular Fermi surface and full spin and valley polarizations (see Extended Data Figure 6 for details). Although at 480 mK the QM-UM phase boundary intercepts the SC1 region, the same phase boundary gradually shifts to lower $n_e$ and eventually enclose the entire SC1 region into the QM phase. This observation indicates that the SC1 state develops from the spin- and valley-polarized QM parent state.

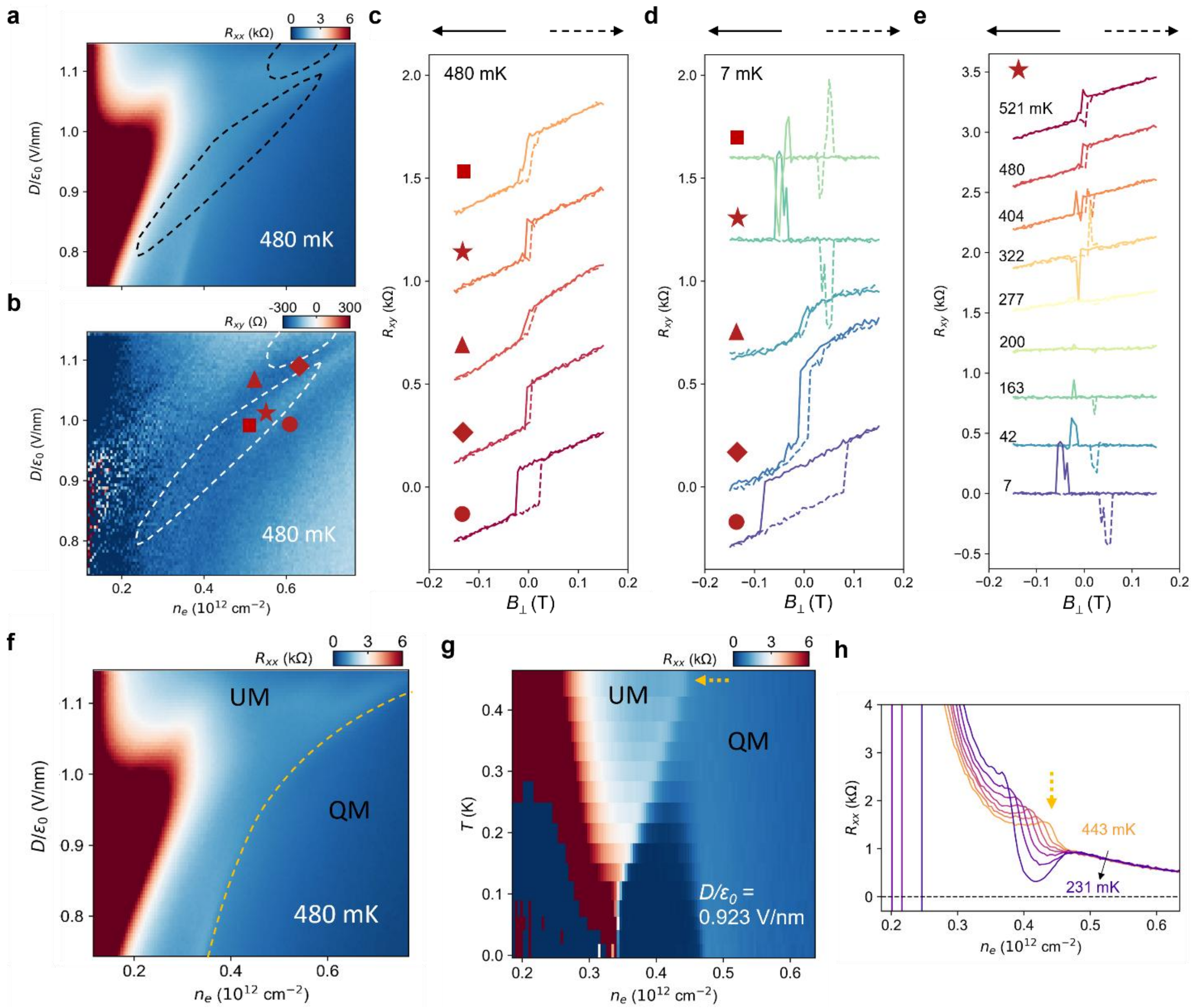

***Figure 4. Temperature-dependent anomalous Hall effects and phase boundary in the tetralayer Device T3.*** ***a&b***, *Symmetrized $R_{xx}$ and anti-symmetrized $R_{xy}$ map at 0.1 T and 480 mK, above the critical temperatures of SC1 and SC2. The dashed curves outline the boundary of SC1 and SC2, inside which clear anomalous Hall signals can be seen in the normal states in* ***b***. ***c&d***, *Magnetic field scans of $R_{xy}$ at the square, star, triangle, diamond and dot positions in* ***b*** *at 480 and 7 mK, respectively. Clear hysteresis can be seen in both the states surrounding SC1, as well as in the SC1 region. Such anomalous Hall signal indicates TRSB due to the orbital degree-of-freedom, which is absent in any previously reported superconductors.* ***e***. *Temperature-dependent $R_{xy}$ hysteresis at the star position. At 277 to 521 mK, non-zero value of $R_{xy}$ at B = 0 T and a linear $R_{xy}$ vs B (the normal Hall signal) can be seen. Below 277 mK, these components disappear due to the superconductivity while clear hysteresis can still be seen.* ***f***. *The same $R_{xx}$ map as in* ***a***, *highlighting (by the orange dashed curve) the phase boundary between the spin- and valley-polarized quarter-metal (QM) and an undetermined metal (UM).* ***g***. *Temperature-dependent $R_{xx}$ line-cut at $D/\varepsilon_0$ = 0.923 V/nm, where the QM-UM phase boundary gradually shifts as T is lowered. The SC1 state develops to the right of the boundary, indicating the QM as the parent state of SC1.* ***h***. *Line-cuts from* ***g***, *showing the QM-UM phase boundary as a kink in $R_{xx}$ which shifts to lower $n_e$ as T is lowered.*

**Strong Coupling of Cooper Pairing**

Lastly, we explore the out-of-plane magnetic-field-dependence of SC1-SC3 in greater details. Figure 5a&b show the $R_{xx}$ in SC1-3 states as a function of $B_{\perp}$ in the tetra- and pentalayer devices, respectively. In both cases, one can see that $R_{xx}$ deviates from zero as $B_{\perp}$ is increased. We define the critical magnetic field $B_{\perp,c}$ as the field when the $R_{xx}$ reaches 10% of the normal state resistance and its uncertainty as the field range between 5% and 15% (see Extended Data Fig. 9), and extract the phenomenological Ginzburg-Landau superconducting coherence length as $\xi_{GL} = \left(\frac{\Phi_0}{2\pi B_{\perp,c}}\right)^{\frac{1}{2}}$, where $\Phi_0 = \frac{h}{2e}$ is the superconducting magnetic flux quantum. Remarkably, $R_{xx}$ in SC1 in the pentalayer device remains within the noise level until $B_{\perp}$ = 1.4 T. Figure 5c summarizes the $\xi_{GL}$ as a function of $n_e$ for SC1-3 at representative displacement fields. As a reference, we plot the inter-electron distance $d_{\text{particle}} = n_e^{-1/2}$ determined by the charge density $n_e$. The coherence length in SC3 is well-above the inter-electron distance. However, the coherence length in SC1 and SC2 are much closer to the latter, especially SC1 in the pentalayer device.

The observations show that SC1 is very unusual that the electrons have a much stronger coupling strength. SC1 is already close to the BCS-BEC crossover[28], although still mainly residing on the BCS side. We note that the critical magnetic field $B_{\perp,c}$ observed in our pentalayer device is higher than any graphene-based superconductors, crystalline or twisted. Compared to the superconducting state in twisted tri-layer graphene[57], the $T_c$ of SC1 in our experiment is more than 10 times lower, but the electron density at which superconductivity is observed is similar, while the critical magnetic field $B_{\perp,c}$ is 2-3 times higher.

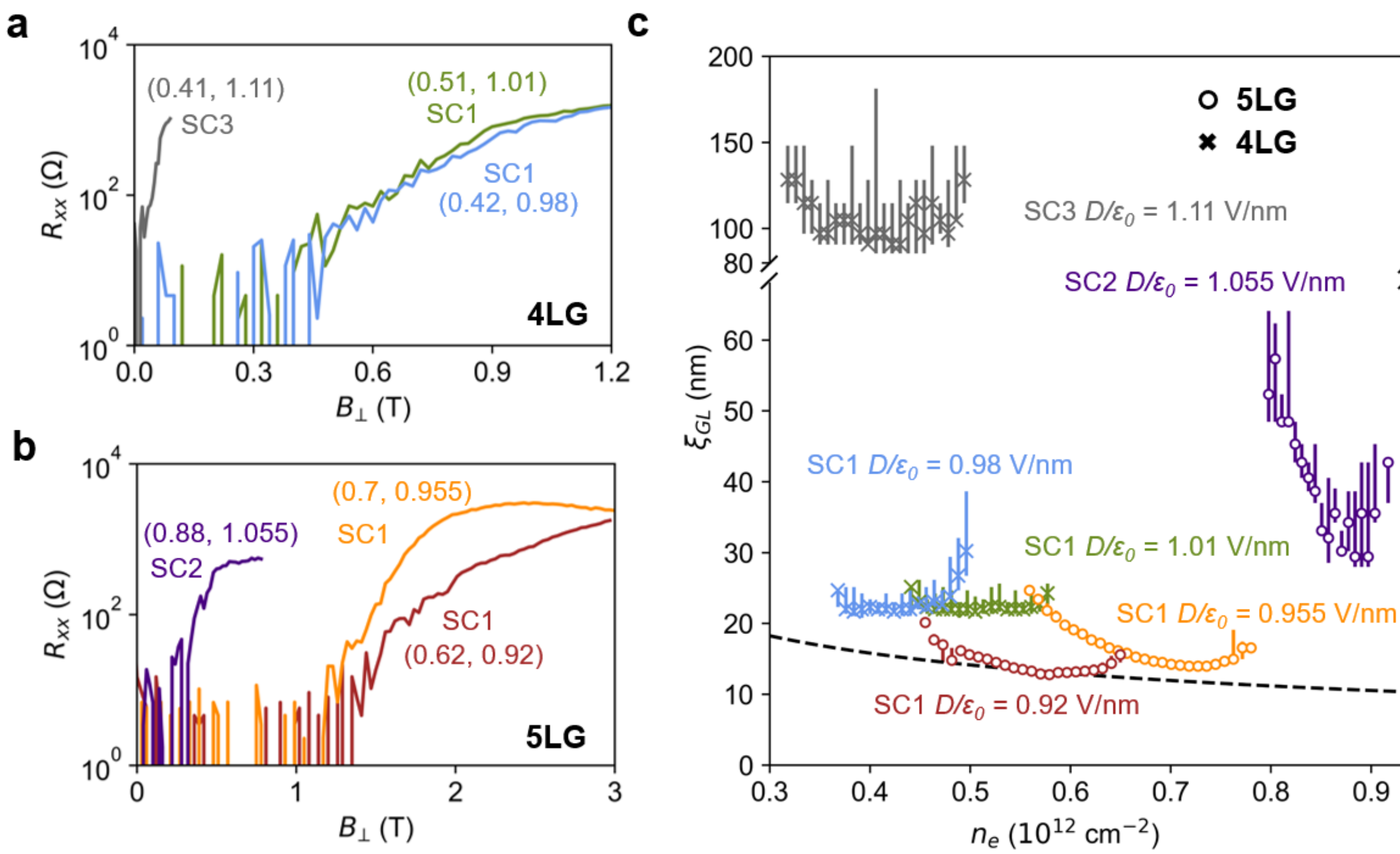

***Figure 5. Superconductivity close to the BCS-BEC crossover. a&b**, Dependence of resistances in SC1-3 on $B_{\perp}$ at 7 mK in the tetralayer Device T3 and pentalayer Device P1, respectively. The curves were taken*

*at (n, D/$\varepsilon_0$) labeled in the figure in the units of ($10^{12}$ $cm^{-2}$, V/nm), respectively.* ***c***, *Coherence length $\xi_{GL}$ as a function of charge density in SC1-3. Here the critical magnetic field and its uncertainties are defined as the field at 10% and field range between 5% and 15% of the normal state resistance, respectively. The dashed lines represent the inter-particle distance derived from the corresponding $n_e$. The $\xi_{GL}$ in SC1 in the pentalayer device is close to the inter-particle distance, indicating strongly coupled Cooper pairing that is close to the BEC-BCS crossover but is still mainly on the BCS side.*

## Discussion

To summarize, we observed two superconducting states SC1 and SC2 that exhibit unusual properties: 1. Magnetic hysteresis and orbital magnetism in the superconducting states; 2. SC1 develops within a spin- and valley-polarized quarter-metal phase, and is robust against the remaining QM state under an out-of-plane magnetic field; 3. the non-zero anomalous Hall signals at zero magnetic field and clear magnetic hysteresis at temperatures above $T_c$. These observations clearly suggest unconventional superconductivity that is distinct from any existing superconductors. These observations suggest spontaneous TRSB at the orbital level in the superconducting states, which is the defining feature of chiral superconductivity[2–4,6].

Microscopically, our observations indicate a spin- and valley-symmetry-broken parent state of superconductivity in SC1, and a valley-symmetry-broken parent state of SC2. In SC1, the parent state is a fully spin- and valley-polarized quarter-metal, which has only one pocket at the Fermi level. In SC2, the parent state is likely a metal state with an annular Fermi surface, which might even have occupations in two different-sized pockets located in opposite valleys (which may have full or partial spin/valley polarization). In the quarter metal case, Cooper pairing occurs within same spin states in a single valley, which must have odd angular momentum due to Pauli exclusion principle, e.g., p-wave or f-wave. Due to the presence of Berry curvature in the valley-polarized state as evidenced by the anomalous Hall effect above $T_c$, we expect the complex-valued chiral order parameter such as p+ip is favored over the real order parameter such as $p_x$. Such chiral superconductors with a single non-degenerate Fermi pocket in two dimensions may be topologically nontrivial and host localized Majorana modes in the vortex core and chiral Majorana fermions at the boundary[2]. We also note that intravalley pairing leads to a large Cooper pair momentum, thus realizing a finite-momentum superconductor[25,26,66–68]. We note that in roughly the same $n_e$-$D$ range hosting SC1-3, tetra- to hexa-layer rhombohedral graphene/hBN moiré superlattice devices show fractional quantum anomalous Hall effects that are hosted by a valley- and spin-polarized topological flat band[58–60].

Our experiment demonstrates a new platform based on simple crystalline graphene for exploring topological superconductivity with local and chiral Majorana zero modes[1–7]. To understand the superconducting ground states that we have observed, future experiments may be performed in several exciting directions: 1. directly probing the TRSB and the orbital magnetism in the superconducting state by using Kerr rotation optical spectroscopy[69] or scanning SQUID[65,70,71]; 2. determining the superconducting gap symmetry by measuring the Fraunhofer pattern of in-plane Josephson junctions[72,73] or Little-Parks effect[74]; 3. characterizing the distribution of supercurrent in magnetic field[75,76] and/or by directly imaging the possible persistent edge current by scanning SQUID[70]; 4. testing quantized thermal conductance of possible Majorana chiral modes on the edges[77]. Our experiment opens up new directions in superconductivity and electron topology physics, and could pave the way to non-abelian-quasi-particle engineering for topologically protected quantum computation applications.

## Methods

## Device fabrication

The graphene, $WSe_2$ (from HQ graphene) and hBN flakes were prepared by mechanical exfoliation onto $SiO_2$/Si substrates. The rhombohedral domains of tetra-layer and penta-layer graphene were identified and confirmed using IR camera[59], near-field infrared microscopy, and Raman spectroscopy and isolated by cutting with a femtosecond laser. The van der Waals heterostructure was made following a dry transfer procedure. We picked up the top hBN, graphite, middle hBN, $WSe_2$ and the tetralayer (pentalayer) graphene using polypropylene carbonate film and landed it on a prepared bottom stack consisting of an hBN and graphite bottom gate. We misaligned the long straight edge of the graphene and hBN flakes to avoid forming a large moiré superlattice. The device was then etched into a multiterminal structure using standard e-beam lithography and reactive-ion etching. We deposited Cr–Au for electrical connections to the source, drain and gate electrodes.

## Transport measurement

The devices were measured mainly in a Bluefors LD250 dilution refrigerator with a lowest electronic temperature of around 40 mK. Stanford Research Systems SR830 lock-in amplifiers at MIT were used to measure the longitudinal and Hall resistance $R_{xx}$ and $R_{xy}$ with an AC frequency at 17.77 Hz. The DC and AC currents are generated by Keysight 33210A function generator through a 100 MΩ resistor. The AC current excitation was limited to be below 0.5 nA. Device T1 was also measured in an Oxford dilution refrigerator at Florida State University. Device P1 was also measured at the University of Basel in a Leiden MNK126-700 dilution refrigerator with a base temperature of ~5mK. An MFLI Zurich Instrument lock-in amplifier (at 17.77 Hz) modulated the AC signal on the DC, followed by a 1 MΩ resistor to fix the current. Basel Precision instrument preamps were used to measure differential currents and voltages. Keithley 2400 source-meters were used to apply top and bottom gate voltages. Top-gate voltage $V_t$ and bottom-gate voltage $V_b$ are swept to adjust doping density $n_e = (C_tV_t + C_bV_b)/e$ and displacement field $D/\varepsilon_0 = (C_tV_t - C_bV_b)/2$, where $C_t$ and $C_b$ are top-gate and bottom-gate capacitance per area calculated from the Landau fan diagram.

## Tight-binding model calculation

The single-particle band structure of the rhombohedral stacked tetralayer graphene is calculated from an effective 8-band Slonczewski-Weiss-McClure type tight-binding model

$$
H = \begin{pmatrix}
u/2 & v_0\pi^\dagger & v_4\pi^\dagger & v_3\pi & 0 & \gamma_2/2 & 0 & 0 \\
v_0\pi & u/2 & \gamma_1 & v_4\pi^\dagger & 0 & 0 & 0 & 0 \\
v_4\pi & \gamma_1 & u/6 & v_0\pi^\dagger & v_4\pi^\dagger & v_3\pi & 0 & \gamma_2/2 \\
v_3\pi^\dagger & v_4\pi & v_0\pi & u/6 & \gamma_1 & v_4\pi^\dagger & 0 & 0 \\
0 & 0 & v_4\pi & \gamma_1 & -u/6 & v_0\pi^\dagger & v_4\pi^\dagger & v_3\pi \\
\gamma_2/2 & 0 & v_3\pi^\dagger & v_4\pi & v_0\pi & -u/6 & \gamma_1 & v_4\pi^\dagger \\
0 & 0 & 0 & 0 & v_4\pi & \gamma_1 & -u/2 & v_0\pi^\dagger \\
0 & 0 & \gamma_2/2 & 0 & v_3\pi^\dagger & v_4\pi & v_0\pi & -u/2
\end{pmatrix}
$$

in the basis of (A1, B1, A2, B2, A3, B3, A4, B4), like the trilayer case[78,79]. Here $v_i = \sqrt{3}a_0\gamma_i/2\hbar$ and $a_0$ = 0.246 nm. The parameters we used are: $\gamma_0$=3.25eV, $\gamma_1$=0.400 eV, $\gamma_2$=- -0.0166eV, $\gamma_3$=-0.293 eV, $\gamma_4$=-0.144 eV. A perpendicular displacement field can introduce a screened potential difference between the top and bottom layers, denoted by Δ. The band structure for rhombohedral pentalayer graphene is calculated using the same parameters using a 10-band model.

The estimation of effective mass in this case is complex due to the trigonally-warped non-parabolic band structure. The effective mass is highly dependent on the density and electric field. We define an averaged effective mass by calculating the density and average kinetic energy[80]

$$
n = \int_{E_m}^{E_F} \frac{d^2\boldsymbol{k}}{(2\pi)^2}\, , W = \frac{1}{n}\int_{E_m}^{E_F} \frac{d^2\boldsymbol{k}}{(2\pi)^2}(E(\boldsymbol{k}) - E_m)
$$

where $E_F$ and $E_m$ denotes the Fermi energy and the conduction band minimum respectively. $E(\boldsymbol{k})$ is the band energy at momentum $\boldsymbol{k}$. Then we compare this to a parabolic band with the same density $n$ and same average kinetic energy $W$ and get the effective mass. We plot the effective mass $m^*$ and Fermi energy $E_F$ as a function of density when $\Delta$ = 90 meV (Extended Data Fig. 12a) and when $n_e = 0.5 * 10^{12}$ cm$^{-2}$ (Extended Data Fig. 12b) near which superconductivity appears. We also plot the effective mass $m^*$ and Fermi energy $E_F$ as a function of density when $\Delta$ = 63 meV (Extended Data Fig. 12c) and when $n_e = 0.6 * 10^{12}$ cm$^{-2}$ (Extended Data Fig. 12d) The calculation assumes there is only one single-valley polarized band, suggested by the experiment.

**Device T1, T2 and P2**

Device T2 has a monolayer $WSe_2$ on top of the tetralayer graphene. Device T1 has a bilayer $WSe_2$ beneath the tetralayer graphene. Device P2 is a bare pentalayer graphene without $WSe_2$. Due to the contact geometry, we can reliably measure the superconducting phases only when the electrons in the conduction band are pushed towards the $WSe_2$ in Device T1, and when electrons in the conduction band are pushed away from $WSe_2$ in Device T2.

The general phase diagram of Device T1 and T2 are similar to that of Device T3. In Device T1, SC1, SC2 and SC3 are observed (Extended Data Fig. 7&8). At $B = 0$ T, both SC1 and SC2 show fluctuations when scanning the gate voltages while SC3 does not. At $B_{\perp} = 0.1$ T, SC3 is destroyed while SC1 and SC2 remains. Magnetic field scans reveal anomalous Hall signals surrounding SC1 (Extended Data Fig. 7). There is also magnetic hysteresis inside SC1. SC1 survives up to $B_{\perp} \sim 0.8$ T and the phase boundary between SC1 and the higher-density-quarter-metal (QM) remains unchanged or even slightly leans towards the QM (Extended Data Fig. 8). SC2 survives up to ~ 0.4 T (Extended Data Fig. 8).

In Device T2, we observed SC1, SC2 and SC3, as well as an additional SC4 (Extended Data Fig. 1). The phase boundary of quarter metal shifts to lower density as the temperature decreases and SC1 emerges from the QM. Such behavior was observed in all three devices (Extended Data Fig. 11).

Although sharing similar qualitative behaviors, the three devices are quantitatively different. For example, the $T_{\mathrm{BKT,\ SC1}}$ for Device T1, T2 and T3 are 160 mK, 210 mK and 300 mK respectively. The difference could originate from the existence of $WSe_2$ and also the device quality variations.

The general phase diagram and behaviors of Device P1 (Extended Data Fig. 9) and P2 (Extended Data Fig. 10) are similar.

**Fluctuations in resistance maps and time domain**

When measuring $R_{xx}$ maps at close to zero magnetic field, we often observe fluctuations in the SC1, SC2 and the neighboring QM states. This is a universal observation (see Fig. 1 for T3, Extended Data Fig. 1 for T2, Extended Data Fig. 7 for T1). The frequency of such fluctuations, however, depends on the details of the specific device and measurement, such as the coercive magnetic field (less when the coercive field is bigger) and the cooling history (less when field-cooled). When fixing the $n_e$ and $D$, it is also possible to see fluctuations of $R_{xx}$ as a function of time, such as shown in Extended Data Fig. 1j. The fluctuations in the QM state we observed has also been observed in previous experiments in rhombohedral trilayer graphene, and was attributed to flipping of the valley-polarization and orbital magnetism[37]. The fluctuations we observed in the SC1 and SC2 states, however, have not been reported in any superconductors. We believe the origin of these fluctuations in the superconducting states is also the flipping of valley-polarization and orbital magnetism, supported by various data in main text and Extended Data Figures.

**Quantum oscillations and fermiology of the neighboring states of SC1 and SC2**

While the spin- and valley-polarized quarter metal are clearly established by the quantum oscillation data and the valley-orbital-magnetic hysteresis, the Fermi surface topology in the UM state in Fig. 4f is much less clear. This can be seen from Fig. 2 and Extended Data Fig. 9, where no clear quantum oscillations can be observed in the region to the lower density side of the QM-UM phase boundary.

Extended Data Fig. 6b&c show the Landau fan at $D/\varepsilon_0$ = 1.123 V/nm and the corresponding FFT spectra. Quantum oscillations can be seen starting at ~1.5 T in the former panel, and a diagonal feature can be seen in the latter panel. The diagonal feature is similar to that observed in the annular Fermi-surfaced metal state in rhombohedral tri-layer graphene[48], which has a frequency above 1. The corresponding low-frequency feature observed in tri-layer graphene, however, is missing from our data. Admittedly, the low-frequency component of FFT is usually more difficult to extract. This is especially true in our case, due to the large effective mass in the flat conduction band. Based on these observations, we can only speculate the UM state to be possibly a spin- and valley-polarized quarter-metal with an annular Fermi surface. This undetermined nature of the UM state (which is to the lower-density-side of the QM-UM phase boundary), however, does not affect visualizing the temperature dependence of phase evolutions and our conclusion of SC1 stemming from a spin- and valley-polarized QM parent state.

**Extraction of coherence length**

We define the critical magnetic field $B_{\perp,c}$ as the field when the $R_{xx}$ reaches 10% of the normal state resistance and its uncertainty as the field range between 5% and 15% (see Extended Data Fig. 9), and extract the phenomenological Ginzburg-Landau superconducting coherence length as $\xi_{GL} = \left(\frac{\Phi_0}{2\pi B_{\perp,c}}\right)^{\frac{1}{2}}$, where $\Phi_0 = \frac{h}{2e}$ is the superconducting magnetic flux quantum. We note that our coherence length is extracted directly from the critical magnetic field, instead of using the Ginzburg–Landau relation $T_c/T_{c0}=1-(2\pi\xi_{GL}^2/\Phi_0)B_\perp$ (where $T_{c0}$ is the mean-field critical temperature at zero magnetic field) and performing a linear fitting near $T_c$. An analysis of SC1 based on the latter approach will result in an even shorter coherence length and even stronger coupling strength.

**Author Contributions**

L.J. supervised the project. T.H. and Z.L. performed the DC magneto-transport measurement. Z.H., A.C., O.S, H.W. and D.M.Z. performed some of the in-plane field measurements (Basel). T.H., L.S., Z.W., W.X., Y.Y. and S.Y. fabricated the devices. J.Y., J.S., Z.L. and T.H. helped with installing and testing the dilution refrigerator. T.H. and M.Z. performed the band structure calculation. H.L., G.S., Z.H. and P.X. helped with part of measurements on Device T1. K.W. and T.T. grew hBN crystals. L.F. contributed to data analysis. All authors discussed the results and wrote the paper.

**Competing Interests** The authors declare no competing interests.

**Data availability** The data that support the findings of this study are available from the corresponding authors upon reasonable request.

## Extended Data Figures

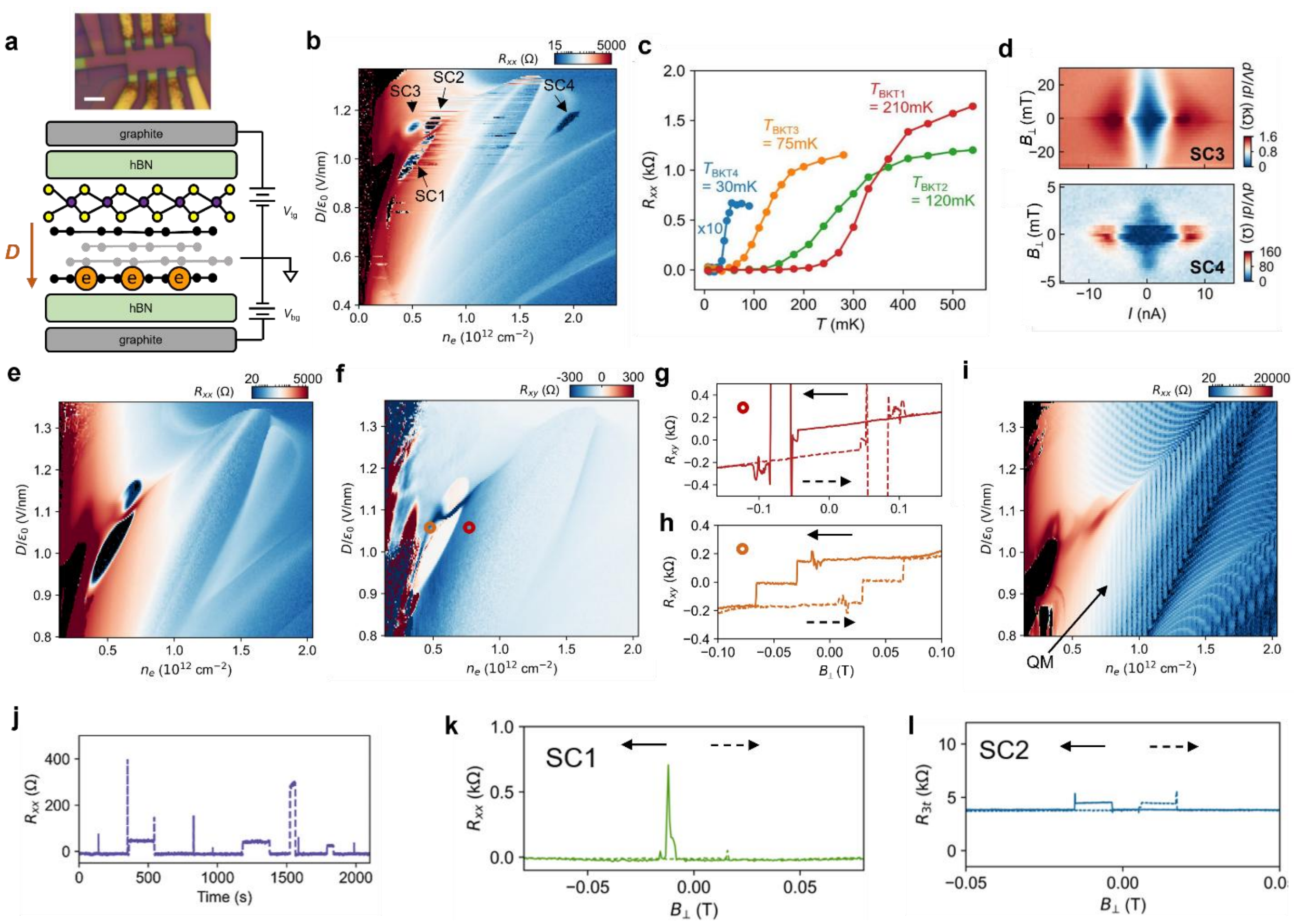

***Extended Data Figure 1. Superconductivity in rhombohedral tetra-layer graphene Device T2. a**, Optical micrograph (Scale bar: 3μm.) and illustration of the structure of rhombohedral tetra-layer graphene,*

*where the electrons are polarized to the layer far away from $WSe_2$.* ***b***, *Four-terminal resistance $R_{xx}$ as a function of $n_e$ and gate displacement field $D/\varepsilon_0$. Four regions show zero $R_{xx}$ (labeled as 'SC1-4' respectively) and superconductivity. SC1 and SC2 regions show fluctuations while SC3 and SC4 are smooth.* ***c***, *Temperature dependence of the four superconducting states, with critical temperatures extracted from the comparison of I-V with the BKT model. See Extended Data Fig. 2.* ***d***, *Differential resistance $dV_{xx}/dI$ as a function of current I and out-of-plane magnetic field $B_\perp$ in the SC3 and SC4 states, respectively. Both states show peaks of dV/dI as a signature of superconductivity at small magnetic fields. The superconductivity is killed below 30 mT, similar to that of most graphene-based superconductors.* ***e&f***, *$R_{xx}$ and $R_{xy}$ maps at 0.1 T, extracted by symmetrizing and anti-symmetrizing the data taken at $B_\perp = \pm 0.1$ T. The fluctuations in SC1, SC2 and neighboring states all disappear. In* ***f***, *SC1 (SC2) is surrounded (neighbored) by states that show anomalous Hall signals. The value of normal Hall signals at the same ne can be seen in the high-D part of the map.* ***g&h***, *Magnetic hysteresis scans of $R_{xy}$ taken at the red and orange circle positions in d, showing loops that are consistent with the anomalous Hall signals in f.* ***i***, *$R_{xx}$ map taken at $B_\perp = 1.5$ T. The period of quantum oscillations indicates a quarter-metal (as labeled by the arrow and 'QM') that neighbors SC1. Combined with the anomalous Hall signals as shown in* ***f***, *this QM is a spin- and valley-polarized phase.* ***j***, *$R_{xx}$ in SC1 (at $n_e = 0.55*10^{12}$ cm$^{-2}$ and $D/\varepsilon_0 = 1.02$ V/nm) as a function of time, featuring fluctuations when gate voltages are fixed.* ***k&l***, *Representative magnetic hysteresis of $R_{xx}$ taken in SC1 and SC2. We note that one of the four terminals was damaged during measurement, resulting in only three-terminal resistance measurement possible.*

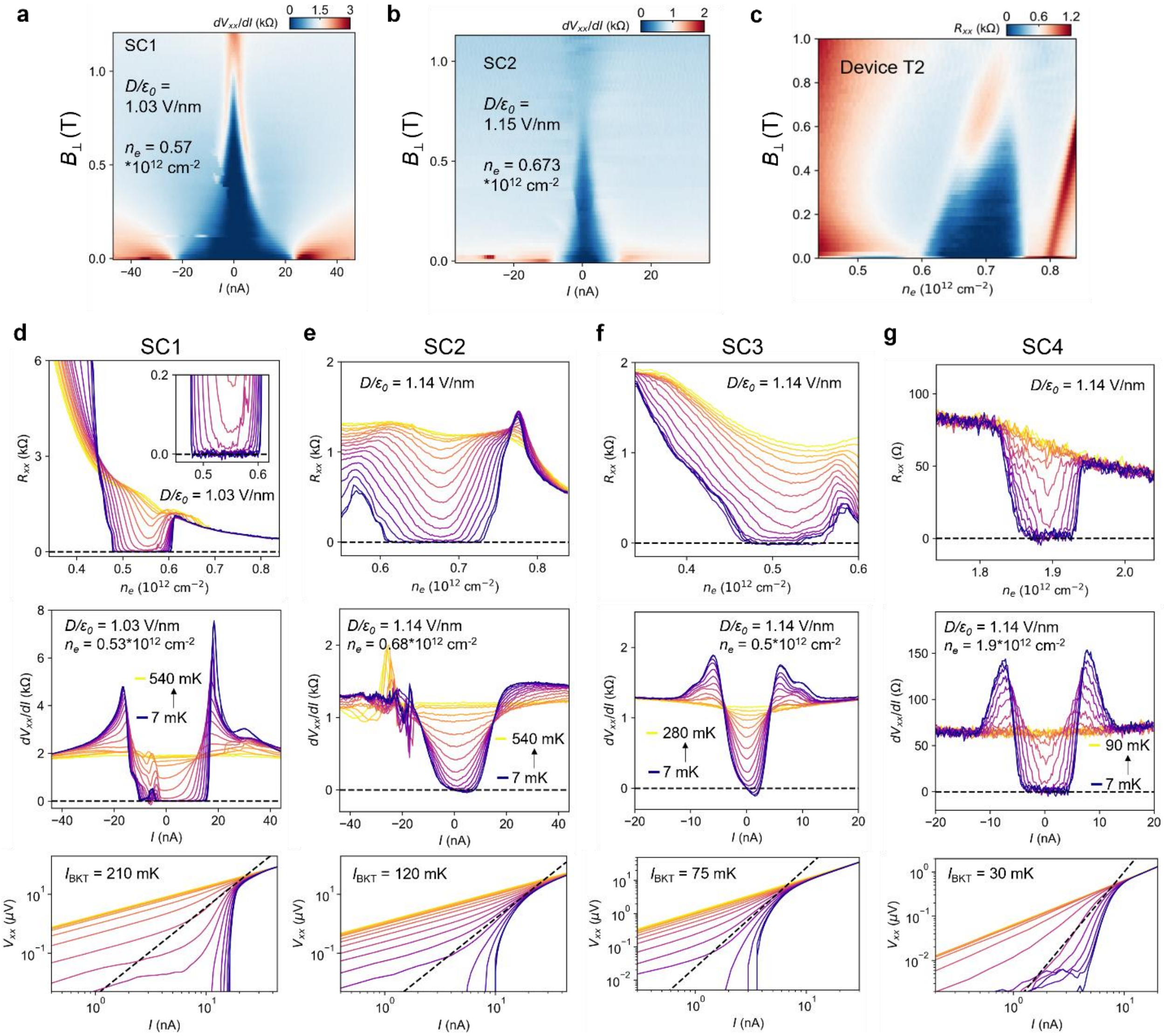

***Extended Data Figure 2. Detailed characterizations of SC1-4 in Device T2. a&b**, Differential resistance $dV_{xx}/dI$ vs $I$ and $B_\perp$ for SC1 and SC2 in Device T2, respectively. The vanishing differential resistance persists to ~ 1 T and ~ 0.6 T in SC1 and SC2, respectively. **c**, B-$n_e$ map at $D/\varepsilon_0$ = 1.14 V/nm. **d-g**, Temperature dependence of longitudinal and differential resistances and BKT fitting for SC1-4. These are taken at representative ($n_e$, D) combinations corresponding to Extended Data Fig. 1c. Panels in the same column correspond to a specific superconducting state. Zero resistance, differential resistance peak at critical current, and the BKT scaling ($V_{xx} \propto I^3$, as indicated by the dashed lines in lower panels) can be seen for all of the four superconducting states.*

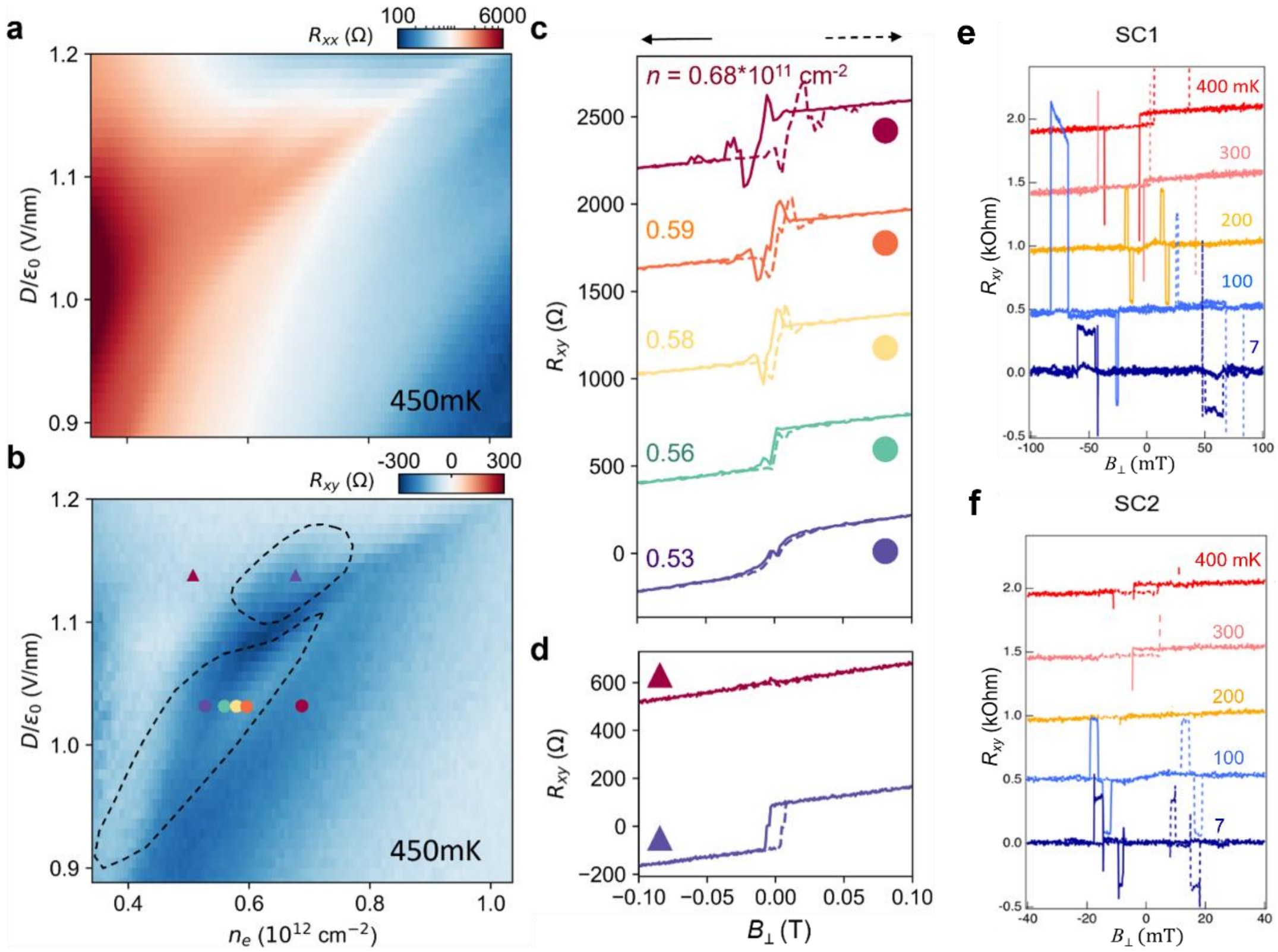

***Extended Data Figure 3. Anomalous Hall effects and TRSB in the normal state of SC1 and SC2 in Device T2. a&b**, Symmetrized $R_{xx}$ and anti-symmetrized $R_{xy}$ map at 0.1 T and 450 mK, above the critical temperatures of SC1 and SC2. The dashed curves in **b** outline the boundary of SC1 and SC2, inside which clear anomalous Hall signals can be seen in the normal states. **c&d**, Magnetic hysteresis scans at the dot and triangle positions in **b**. Clear hysteresis loops can be seen in both the states surrounding SC1, as well as in SC1 and SC2. **e&f**, Temperature-dependent anti-symmetrized $R_{xy}$ hysteresis at a state in SC1 and SC2, respectively. Curves are shifted vertically for clarity.*

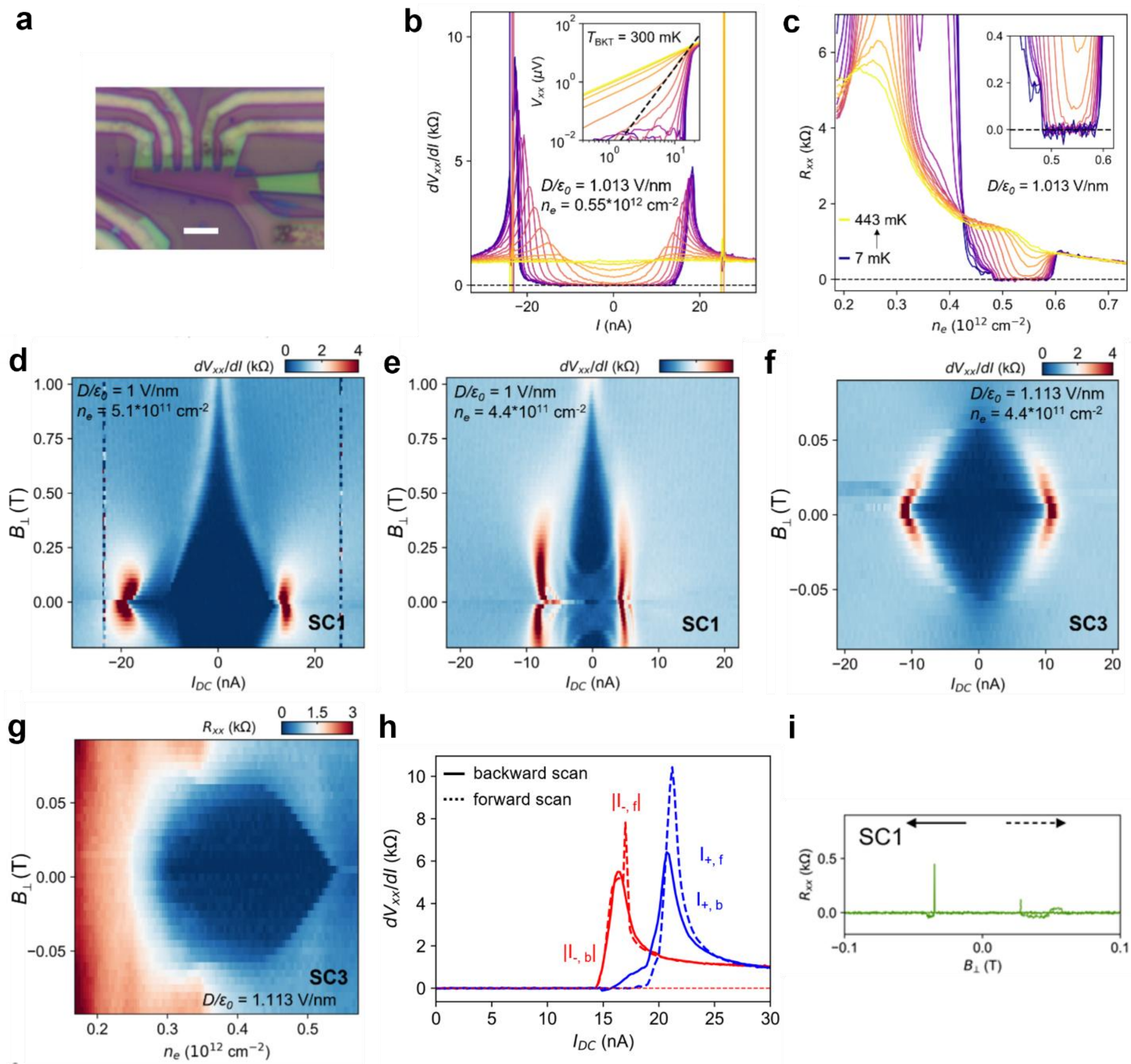

***Extended Data Figure 4. Superconductivities in Device T3. a**, Optical micrograph of the device. Scale bar: 3μm. **b**, Temperature-dependent differential resistance $dV_{xx}/dI$ versus I at a typical ($n_e$, D) inside the SC1 region, featuring zero resistance at low current and a pair of peaks at critical current. **c**, Temperature-dependent $R_{xx}$ at a constant D, featuring a density range of zero resistance that corresponds to SC1. **d-f**, Differential resistance at typical $n_e$-D positions inside SC1 and SC3. The vanishing differential resistance persists to ~ 1 T for SC1, while that of SC3 persists to only ~ 50 mT. **g**, $R_{xx}$ as a function of $n_e$ and $B_\perp$ at $D/\varepsilon_0$ = 1.113 V/nm in SC3. The density range corresponding to SC3 keeps shrinking upon $B_\perp$. **h**, Differential resistance measurement in SC1, showing the superconducting diode effect. **i**, Representative magnetic hysteresis of $R_{xx}$ taken in SC1.*

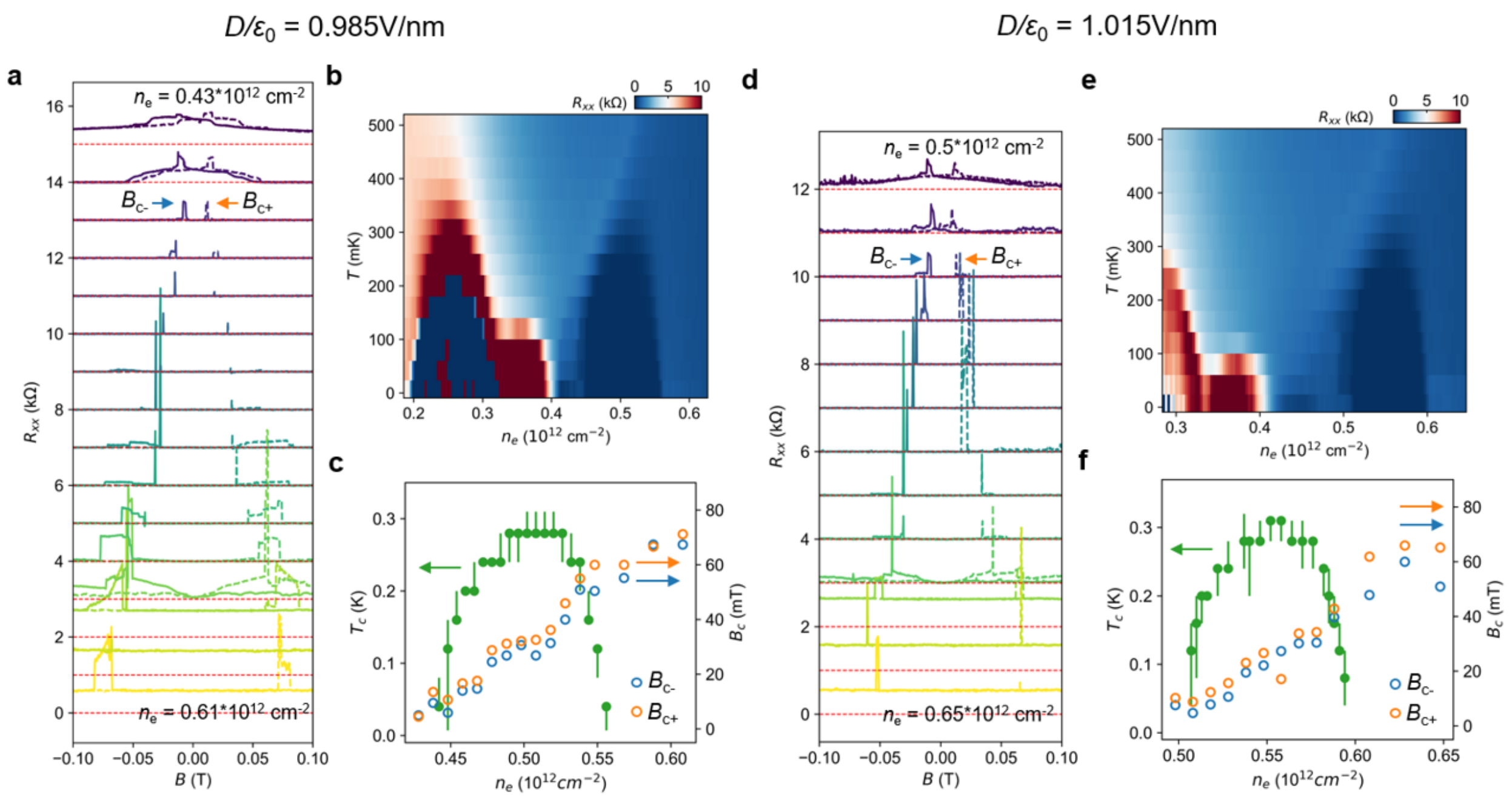

***Extended Data Figure 5. Magnetic hysteresis, coercive field and superconducting critical temperature in SC1 in Device T3. a**, $R_{xx}$ as a function of the out-of-plane magnetic field at different $n_e$ and $D/\varepsilon_0$ = 0.985 V/nm. The curves are shifted vertically for clarity. The dashed horizontal lines indicate the shift of each curve (which corresponds to zero resistance). Orange and blue arrows indicate the coercive fields, which is defined as the closest-to-zero magnetic field where $R_{xx}$ rises rapidly. **b**, Color map of $R_{xx}$ versus T and $n_e$. **c**, Summary of the coercive fields and the superconducting $T_c$ at different $n_e$ and $D/\varepsilon_0$ = 0.985 V/nm. **d-f**, same as **a-c** but for $D/\varepsilon_0$ = 1.015 V/nm.*

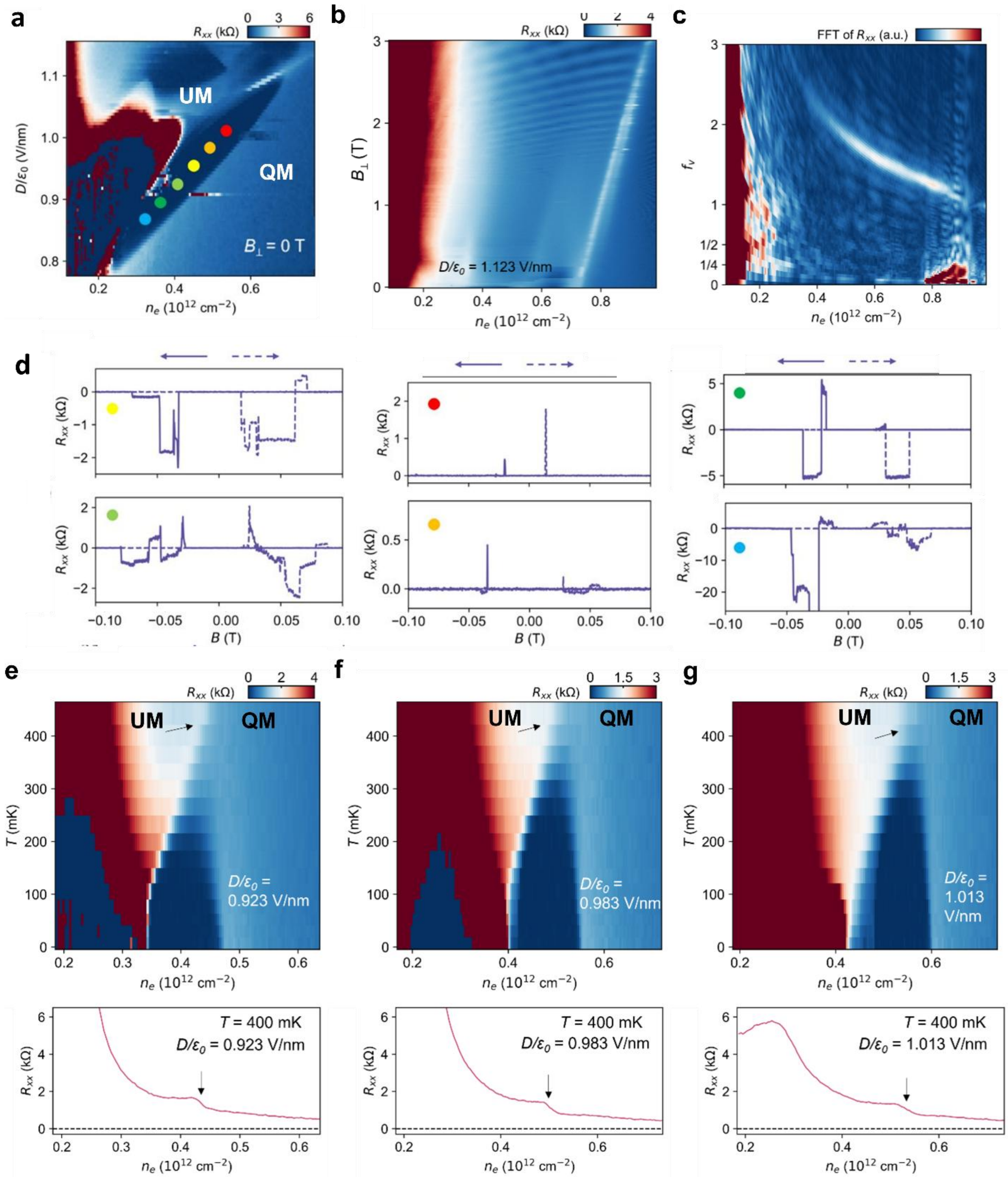

***Extended Data Figure 6. Magnetic hysteresis, quantum oscillations and temperature-dependence of SC1 in Device T3*. *a*, *The $n_e$-$D$ map of $R_{xx}$ taken at zero magnetic field in Device T3.* ***b***, *Landau fan diagram taken at $D/\varepsilon_0$ = 1.123 V/nm, revealing quantum oscillations starting at $B_\perp$ ~ 1.5 T.* ***c***, *Fast Fourier-transform spectra of data in* ***b***. *A diagonal feature above $f_\nu$ = 1 suggests a quarter-metal state with annular Fermi surface. However, the low-frequency component of this annular Fermi-surfaced metal is not clear from the data.* ***d***, *Out-of-plane magnetic field scans of $R_{xx}$ at different ($n_e$, $D$) indicated by the colored dots in* ***a***. *Magnetic hysteresis was observed across a large range of ($n_e$, $D$) parameter space across SC1.* ***e-g***, *Upper*

*panels: $R_{xx}$ as a function of T and $n_e$ at three displacement fields cutting through SC1. In all cases, there is a clear boundary as pointed out by the black arrow at above $T_c$. This boundary shifts to lower $n_e$ values as the temperature is lowered. Superconductivity domes emerge within the phase to the right of this boundary, suggesting this phase to the right (the spin- and valley-polarized quarter-metal) is the parent state of SC1. Lower panels: line-cuts at T = 400 mK from the upper panels, featuring kinks that corresponds to the phase boundary between the spin- and valley-polarized quarter-metal and the metal state at lower densities.*

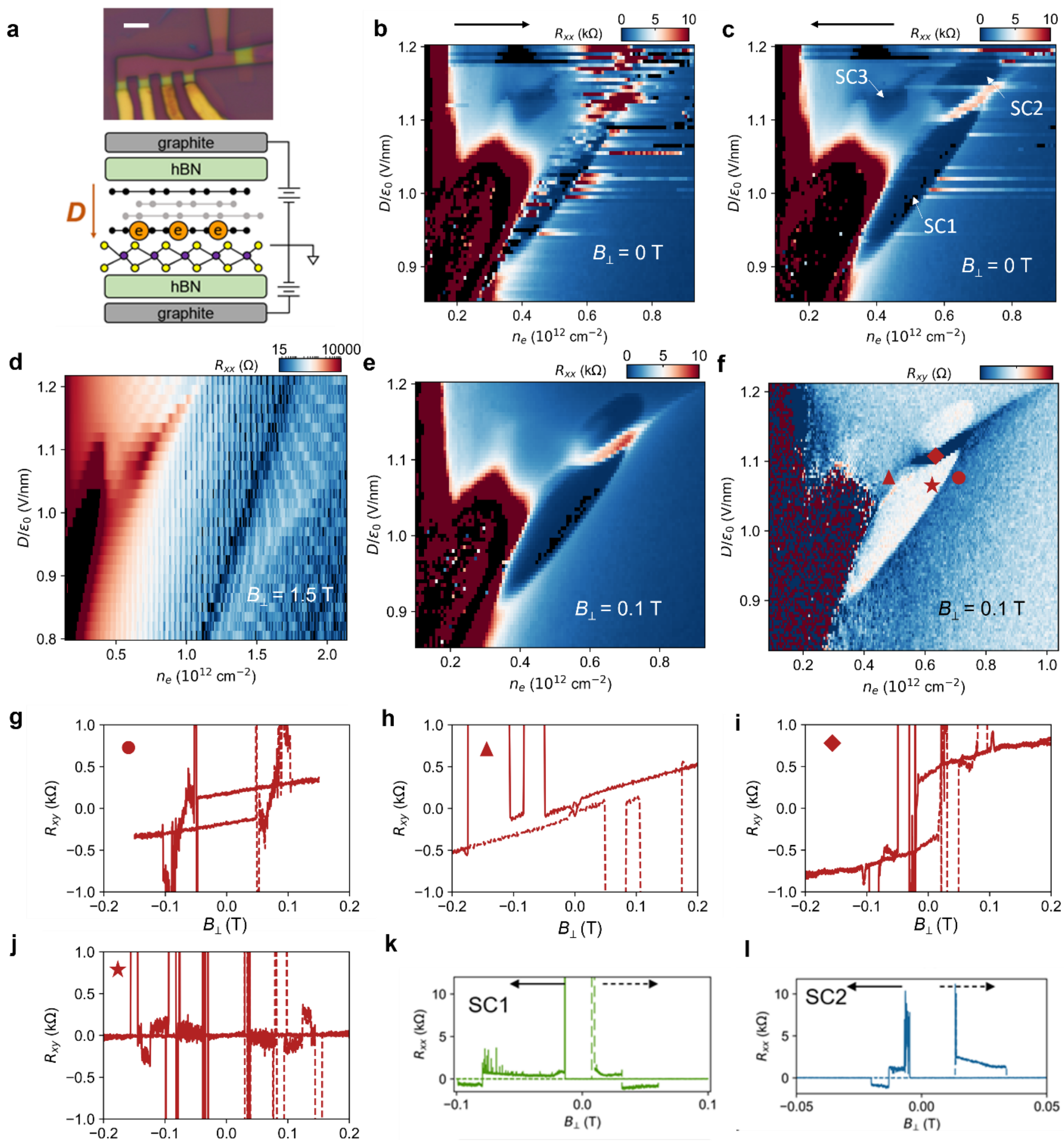

**Extended Data Figure 7. Superconductivities in Device T1. *a***, *Optical micrograph (Scale bar: 3μm) and device configuration, where electrons are polarized to the bottom layer of tetra-layer graphene with WSe2*

*at proximity.* ***b&c****, The $n_e$-D maps of $R_{xx}$ at $B_\perp = 0$ T and base temperature, corresponding to opposite sweeping directions of $n_e$, respectively. Three superconducting regions labeled as SC1-3 similar to in Device T2 and T3 can be seen. Some fluctuations can be seen in SC1, SC2 and the neighboring metallic region.* ***d****, The $n_e$-D map of $R_{xx}$ at $B_\perp = 1.5$ T and base temperature, featuring the quantum oscillations of a quarter-metal to the right of the SC1 region.* ***e&f****, The $n_e$-D map of $R_{xx}$ and $R_{xy}$ at $B_\perp = 0.1$ T and base temperature. The fluctuations and SC3 are both suppressed, similar to those observed in Device T2.* ***g-j****, Magnetic hysteresis scans of $R_{xy}$ taken at the dot, triangle diamond and star positions in* ***f****, showing jumps/loops that are consistent with the anomalous Hall signals in* ***f****.* ***k&l****, Representative magnetic hysteresis of $R_{xx}$ taken in SC1 and SC2.*

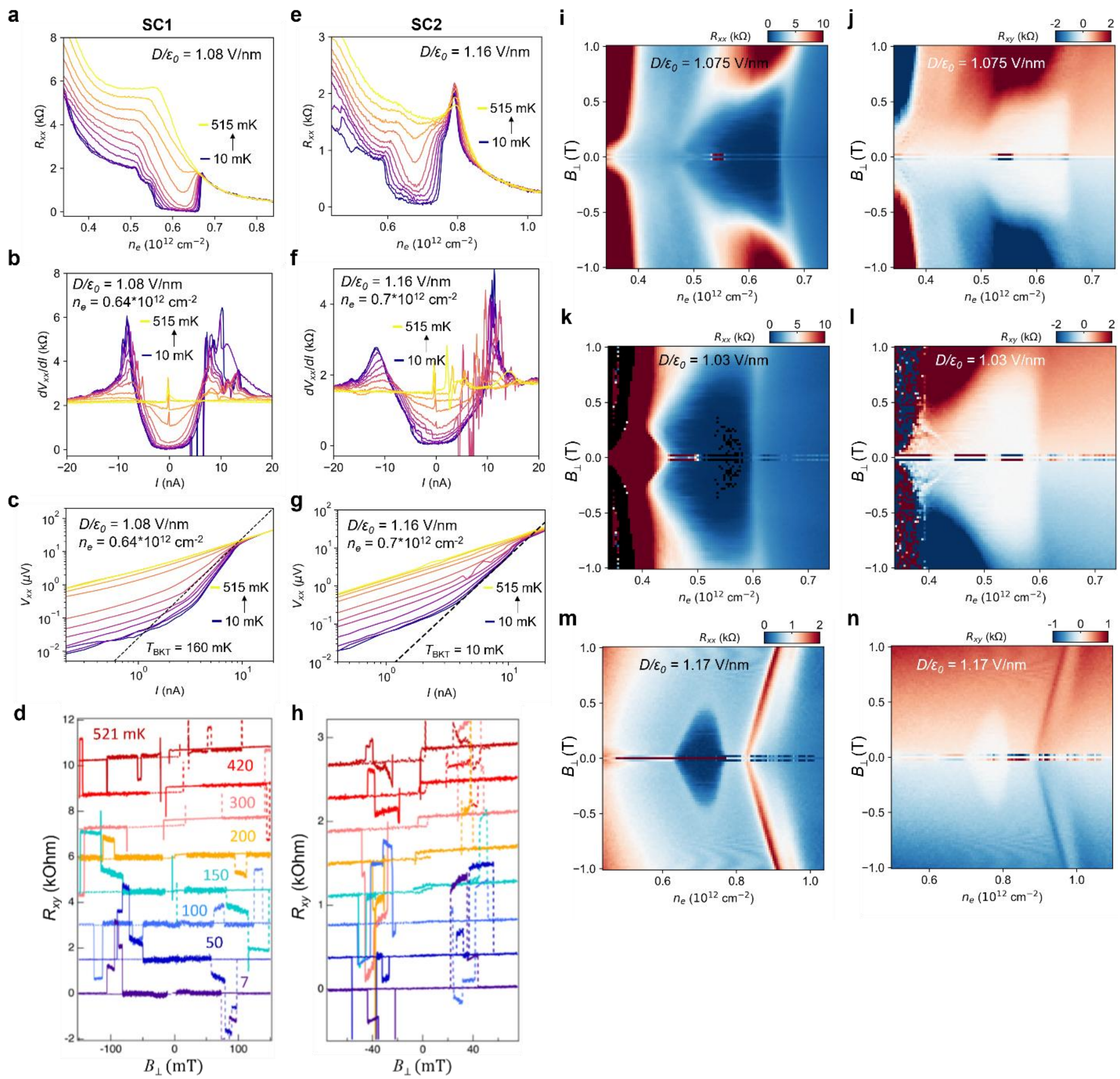

***Extended Data Figure 8. Temperature and magnetic field dependence of superconductivity in Device T1. a-c****, Temperature dependence of $R_{xx}$, the difference resistance $dV_{xx}/dI$ vs I, and the BKT fitting of SC1 respectively.* ***d****, Temperature-dependent anti-symmetrized $R_{xy}$ hysteresis at a state in SC1.* ***e-g****, The*

*temperature dependence of $R_{xx}$, the difference resistance dV/dI vs I, and the BKT fitting of SC2.* ***h****, Temperature-dependent anti-symmetrized $R_{xy}$ hysteresis at a state in SC2.* ***i&j****, $R_{xx}$ and $R_{xy}$ as a function of $n_e$ and $B_\perp$ at $D/\varepsilon_0$ = 1.075 V/nm (corresponding to SC1), respectively. The phase boundary between the quarter-metal and SC1 remains at the same $n_e$, indicating the orbital magnetism is continuous across the boundary and SC1 is orbital magnetic.* ***k&l****, $R_{xx}$ and $R_{xy}$ as a function of $n_e$ and $B_\perp$ at $D/\varepsilon_0$ = 1.03 V/nm (corresponding to SC1), respectively. The phase boundary between the quarter-metal and SC1 remains at the same $n_e$, while the left boundary of SC1 even moves against the neighboring state in magnetic field, confirming the orbital magnetic nature of SC1.* ***m&n****, $R_{xx}$ and $R_{xy}$ as a function of ne and $B_\perp$ at $D/\varepsilon_0$ = 1.17 V/nm (corresponding to SC2), respectively. The phase boundaries between SC2 and neighboring states move towards SC2 under magnetic field.*

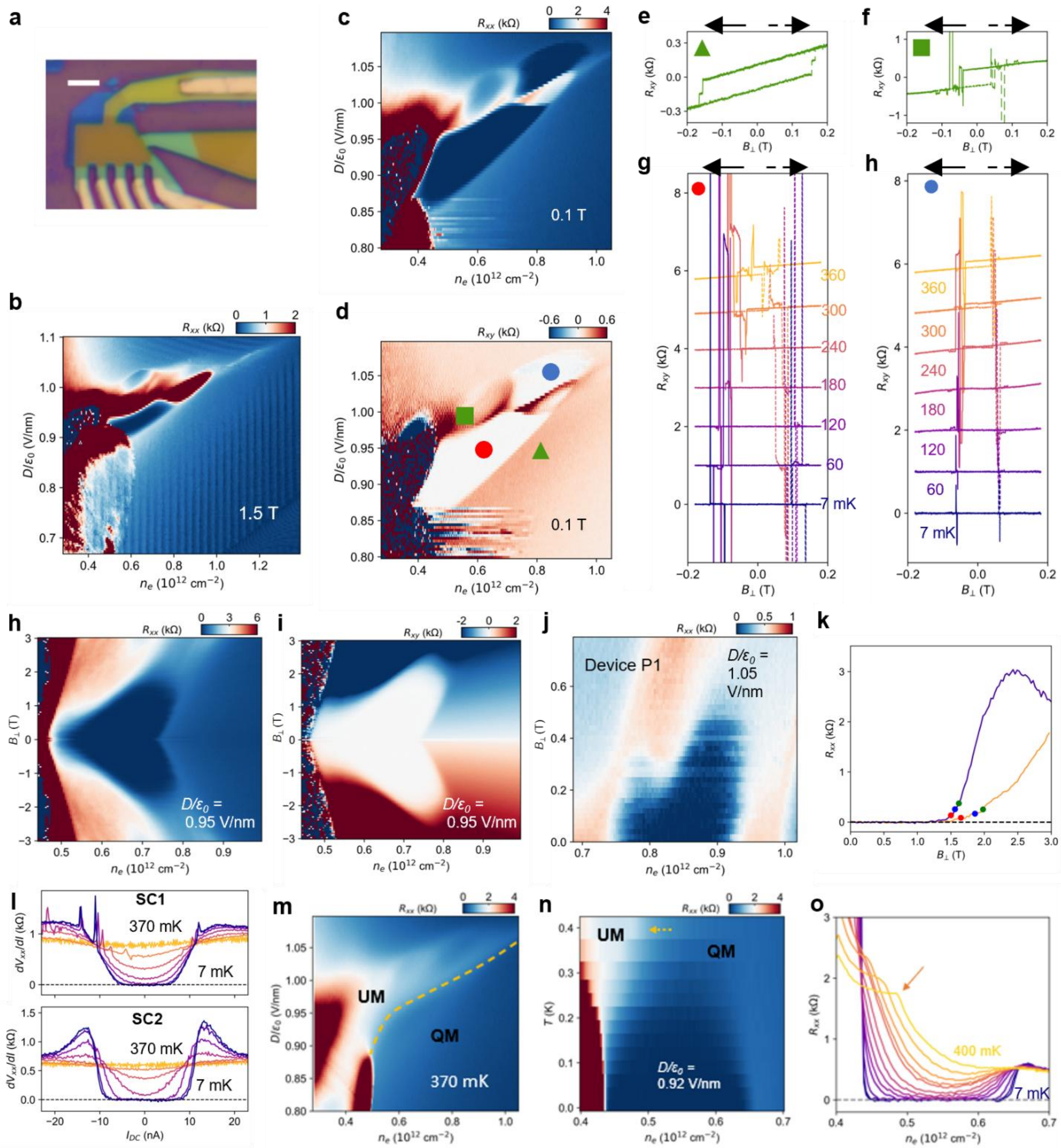

***Extended Data Figure 9. Superconductivities in Device P1. a**, Optical micrograph of the device. Scale bar: 3 $\mu$m. **b**, The $n_e$-$D$ map of $R_{xx}$ at $B_\perp$ = 1.5 T and base temperature, featuring the quantum oscillations corresponding to the quarter-metal state neighboring SC1. **c&d**, The $n_e$-$D$ map of $R_{xx}$ and $R_{xy}$ at $B_\perp$ = 0.1 T and base temperature, respectively. **e&f**, Magnetic hysteresis of $R_{xy}$ at the green triangle and square positions in **d**. **g&h**, Temperature-dependence of anti-symmetrized $R_{xy}$ in SC1 (corresponding to the red dot position in **d**) and SC2 (corresponding to the blue dot position in **d**), respectively. Curves are shifted vertically for clarity. **h&i**, $R_{xx}$ and $R_{xy}$ as a function of $n_e$ and $B_\perp$ at $D/\varepsilon_0$ = 0.955 V/nm, respectively. The phase boundary between the quarter-metal and SC1 shifts to slightly higher density, suggesting the orbital magnetic nature of SC1. **j**, The $n_e$-$B$ map of $R_{xx}$ at $D/\varepsilon_0$ = 1.05 V/nm, cutting through SC2. **k**, Magnetic field-*

*dependence of $R_{xx}$ in two representative states inside SC1. We use 10% (indicated by the blue dots) of the normal state resistance to extract the $T_c$, and 5% (red dots) and 15% (green dots) of the normal state resistance to extract the uncertainty of $T_c$ in Figure 5 of the main text.* ***l**, $dV_{xx}/dI$ versus I in SC1 and SC2 at ($0.61*10^{12}$ $cm^{-2}$, 0.94 V/nm) and ($0.85*10^{12}$ $cm^{-2}$, 1.05 V/nm) respectively, featuring zero-resistance at small current and the resistance spikes at critical current.* ***m**, The $n_e$-D map of $R_{xx}$, highlighting (by the orange dashed curve) the phase boundary between the spin- and valley-polarized quarter-metal (QM) and an undetermined metal (UM).* ***n**. Temperature-dependent $R_{xx}$ line-cut at $D/\varepsilon_0$ = 0.92 V/nm, where the QM-UM phase boundary (indicated by orange dashed arrow) gradually shifts as T is lowered. The SC1 state develops to the right of the boundary, indicating the QM as the parent state of SC1.* ***o**. Line-cuts from **n**, showing the QM-UM phase boundary as a kink (orange arrow) in $R_{xx}$ which shifts to lower $n_e$ as T is lowered.*

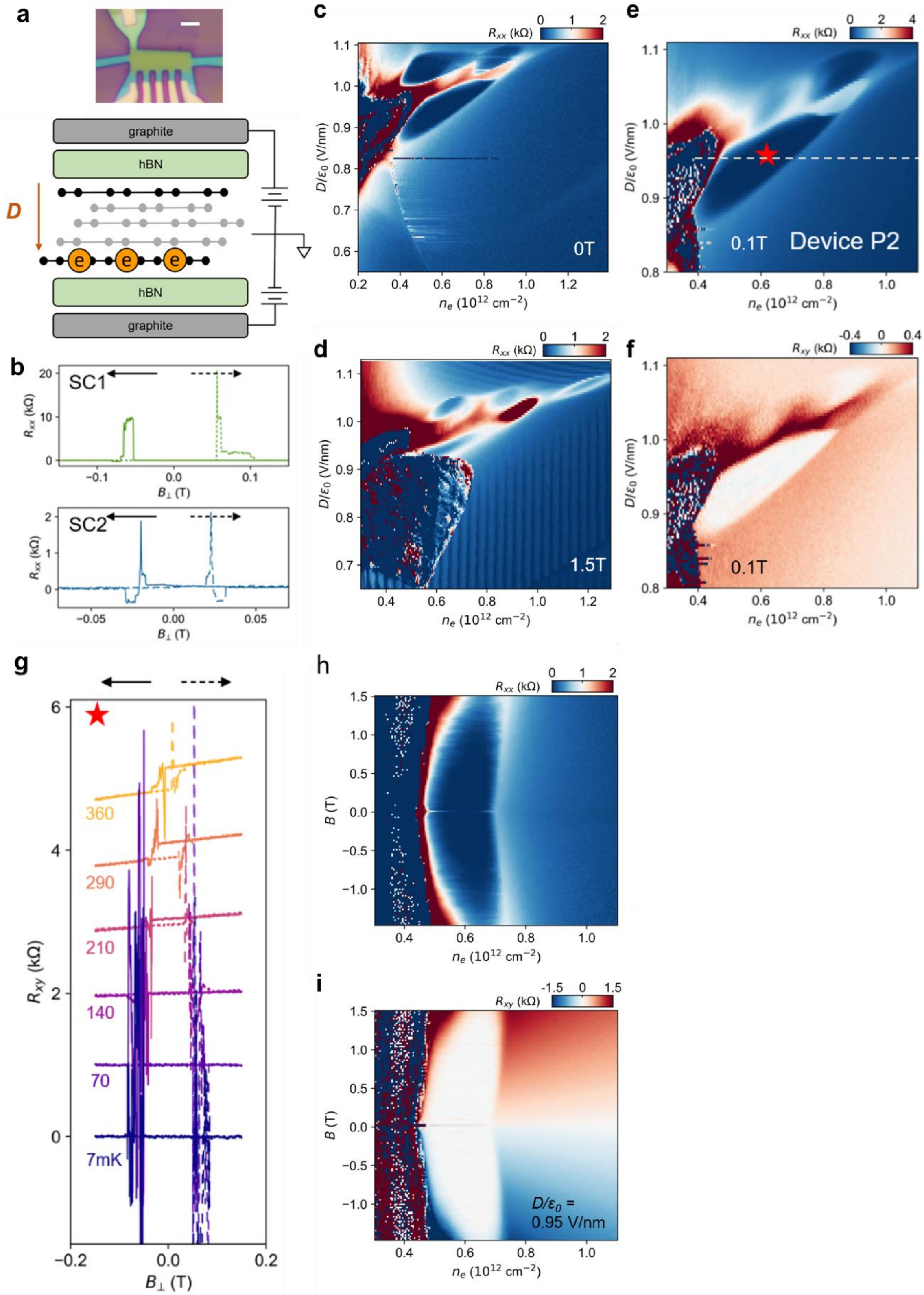

***Extended Data Figure 10. Superconductivities in Device P2. a****, Optical micrograph (scale bar: 3 μm) and illustration of the device configuration.* ***b****, Magnetic hysteresis in SC1 and SC2 at base temperature,*

*respectively.* ***c***, *The $n_e$-D map of $R_{xx}$ at zero magnetic field, featuring SC1-SC3.* ***d***, *The $n_e$-D map of $R_{xx}$ at $B_\perp = 1.5$ T, featuring the quarter-metal state to the higher density side of SC1.* ***e&f***, *The $n_e$-D map of $R_{xx}$ and $R_{xy}$ at $B_\perp = 0.1$ T and base temperature.* ***g***, *Temperature-dependent magnetic hysteresis of $R_{xy}$ at the 'star' position in* ***e***. *Curves are shifted vertically for clarity.* ***h&i***, *$R_{xx}$ and $R_{xy}$ as a function of $n_e$ and $B_\perp$ along the dashed line in* ***e***, *respectively. The phase boundary between the quarter-metal and SC1 shifts to slightly higher density, suggesting the orbital magnetic nature of SC1.*

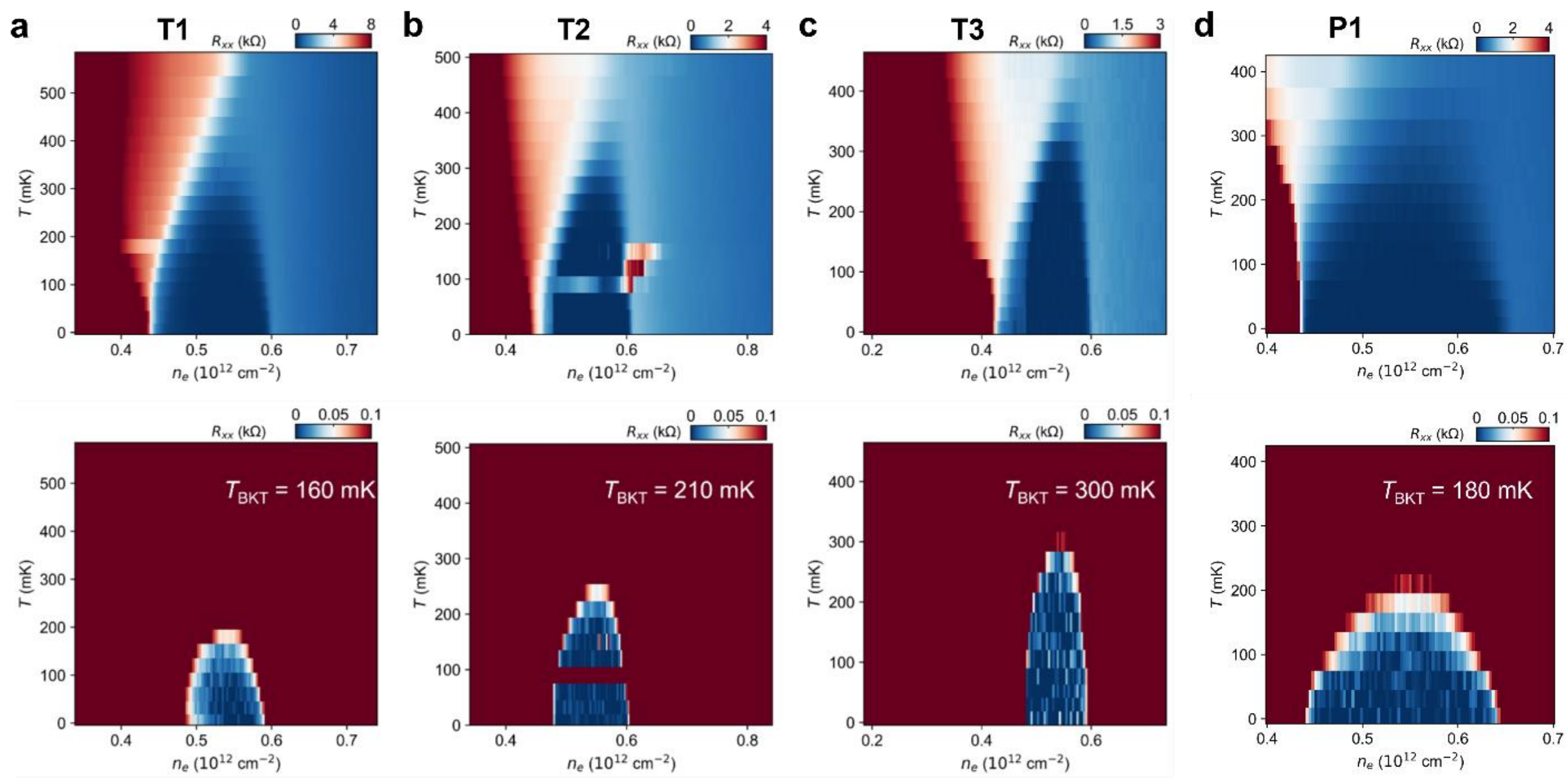

***Extended Data Figure 11. Comparison between the highest superconducting transition temperatures of SC1 in Device T1-T3 and P1. a-d***, *Upper panels: $R_{xx}$ as a function of temperature and charge density at a constant D that corresponds to highest $T_c$, in four devices respectively. Lower panels: the same plots as in upper panels with a small unified color scale for a fair comparison. The BKT fitting reveals an increase of $T_{BKT}$ from Device T1 to T3, corresponding to a weakening of spin-orbit-coupling effect.*

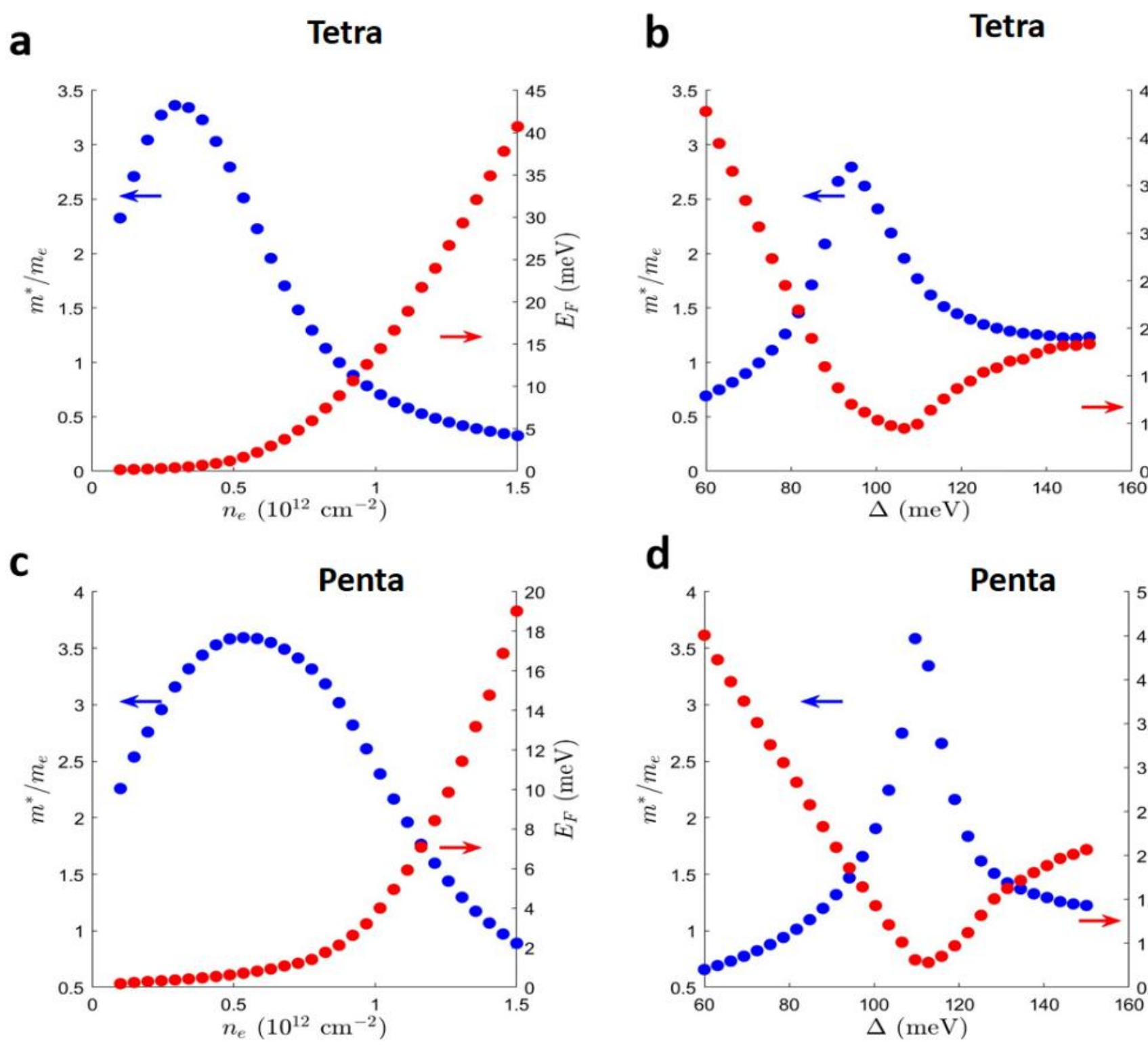

***Extended Data Figure 12. Calculation of the effective mass and Fermi energy in tetra- and penta-layer rhombohedral graphene. a****, Calculation at a fixed potential difference between the top-most and bottom-most layers $\Delta$ = 90meV in tetralayer graphene.* ***b****, Calculation at a fixed charge density $n_e = 0.5*10^{12}$ cm$^{-2}$ in tetralayer graphene.* ***c****, Calculation at a fixed potential difference $\Delta$ = 110 meV in pentalayer graphene.* ***d****, Calculation at a fixed charge density $n_e = 0.6*10^{12}$ cm$^{-2}$ in pentalayer graphene.*